\documentclass[11pt,letterpaper,reqno]{amsart}
\usepackage{graphicx,booktabs,enumitem,xcolor}
\usepackage[T1]{fontenc}
\usepackage{amsfonts,amsmath,amssymb,amsthm,bm,mathrsfs}
\usepackage{tikz}
\usetikzlibrary{arrows.meta,positioning,shapes.misc}
\usepackage[colorlinks=true,linkcolor=blue!55!black,citecolor=blue!55!black,urlcolor=blue!55!black]{hyperref}

\numberwithin{equation}{section}
\numberwithin{figure}{section}
\DeclareMathOperator*{\esssup}{ess\,sup}
\DeclareMathOperator*{\essinf}{ess\,inf}
\newtheorem{theorem}{Theorem}[section]
\newtheorem{proposition}{Proposition}[section]
\newtheorem{lemma}{Lemma}[section]
\newtheorem{corollary}{Corollary}[section]
\theoremstyle{definition}
\newtheorem{assumption}{Assumption}[section]

\theoremstyle{remark}
\newtheorem{remark}{Remark}[section]

\newcommand{\E}{\mathbb{E}}
\newcommand{\R}{\mathbb{R}}
\renewcommand{\P}{\mathbb{P}}
\newcommand{\F}{\mathcal{F}}
\newcommand{\cT}{\mathcal{T}}
\newcommand{\cC}{\mathcal{C}}
\newcommand{\cS}{\mathcal{S}}
\newcommand{\cL}{\mathcal{L}}
\newcommand{\1}{\mathbf{1}}
\newcommand{\eps}{\varepsilon}
\newcommand{\dd}{\,\mathrm{d}}

\title[The end-of-month option and the Gittins index of holding]{The end-of-month option in futures contracts\\ and the Gittins index of holding the underlying}
\author{Erhan Bayraktar}
\address{Department of Mathematics, University of Michigan, Ann Arbor, MI 48109}
\email{erhan@umich.edu}
\thanks{\emph{Dedicated to the memory of Tomas Bj\"ork.} Tomas told the author about the problem studied here during a visit to Ann Arbor in 2007. The author recently returned to the problem after learning that it remained open. Draft of \today.}
\thanks{The author is partially supported by the National Science Foundation under grant DMS-2602036 and by the Susan M. Smith Chair.}
\date{}

\begin{document}

\begin{abstract}
The end-of-month option allows the short in a futures contract to choose when to deliver during a window in which the delivery price is fixed. We characterize optimal delivery through a universal index obtained from the Bank--El Karoui representation theorem. In the semimartingale setting, this is the Gittins index of holding the underlying and collecting its convenience yield, and the futures price is the retirement reward for which the value of this project equals the spot price. This characterization yields price bounds and a criterion for immediate delivery. Dinkelbach's iteration reduces the price computation to a sequence of standard optimal stopping problems, with an explicit contraction bound. We apply the theory to geometric Brownian motion and the Gibson--Schwartz model with stochastic convenience yield. For geometric Brownian motion, we characterize when the futures price equals spot and derive a square-root boundary asymptotic near expiry. For the Gibson--Schwartz model, we establish monotonicity in the yield and characterize the critical yield below which the price equals spot. Numerical results illustrate the price iteration and delivery regions.
\end{abstract}

\maketitle

\section{Introduction}

Many futures contracts permit a choice of delivery date. We study the \emph{end-of-month option}, under which the delivery price has been fixed and the short chooses when to deliver a specified underlying during the remaining delivery window. Let $X$ be the price of the underlying and let $[0,T]$ be the delivery window. Delivery at the stopping time $\tau$ against the fixed price $m$ is worth $m-X_\tau$ to the short, so the short's problem at the price $m$ is
\begin{equation}\label{eq:Vintro}
V(m):=\sup_{\tau\in\cT[0,T]}\E\big[e^{-r\tau}(m-X_\tau)\big],
\end{equation}
and the contract is worth zero at the start of the window exactly when
\begin{equation}\label{eq:zero}
V(m^\ast)=0 .
\end{equation}
Since $V$ is strictly increasing, \eqref{eq:zero} pins the price down as
\begin{equation}\label{eq:ratio}
m^\ast=\inf_{\tau\in\cT[0,T]}\frac{\E[e^{-r\tau}X_\tau]}{\E[e^{-r\tau}]} .
\end{equation}
Biagini and Bj\"ork \cite[pp.~267--268]{BiaginiBjork07} solved the timing option of the delivery month in this journal and distinguished it from the end-of-month option, under which delivery occurs after the futures price has been fixed. Tomas Bj\"ork told the author about it during a visit to Ann Arbor in 2007, in the form \eqref{eq:ratio}: the futures price is a ratio of two expectations, each over the same stopping time, and one wants to know the price and the delivery strategy. A ratio of expected accumulated reward to expected discounted time also appears in the forward-induction description of the Gittins index for a diffusion in \cite{Karatzas84}; this suggests looking for a bandit behind the contract. Lindensj\"o \cite{Lindensjo16} gave the definition and the formula \eqref{eq:ratio} rigorously in a general incomplete market with an adapted interest rate, obtained two explicit cases (if the discounted underlying is a submartingale, deliver at once; with discrete dividends, deliver at activation or at dividend dates), characterized the futures price process when the end-of-month option is combined with the timing option of \cite{BiaginiBjork07}, and wrote that ``there is to our knowledge no general way of directly calculating'' $m^\ast$.

This paper characterizes the optimal delivery strategy through a universal index and reduces the endogenous price problem \eqref{eq:zero} to a provably convergent sequence of standard optimal stopping problems, which we implement in specific models. The index comes from the representation theorem of Bank and El Karoui \cite{BankElKaroui04}, applied to the \emph{excess process} $Z_t=e^{-rt}X_t-\E[e^{-rT}X_T\,|\,\F_t]$, which vanishes at $T$. For each fixed price $m$, the resulting signal $K$ gives an optimal delivery rule: deliver at the first time $K_t\le m$ before the fixed expiry $T$, or at $T$ if no such time occurs (Theorem~\ref{thm:signal}). The price $m^\ast$ is then the unique root of an explicit equation in the running infimum of $K$ (see~\eqref{eq:fixed}). Moreover, $\E[e^{-r\tau^m}]$, where $\tau^m$ denotes this delivery time, lies between the one-sided derivatives of $V$ at $m$, so that $V'(m)=\E[e^{-r\tau^m}]$ wherever $V$ is differentiable.

That stopping problems with a running reward are multi-armed bandits in the sense of Gittins and Whittle, and that their index is the Bank--El Karoui signal, is the unifying view of Bank and F\"ollmer \cite{BankFollmer03}; what is specific to the futures contract is the bandit behind it, identified by the compensator of the discounted underlying. Writing $e^{-rt}X_t=X_0+M_t-A_t$ with $M$ a martingale and $A$ the compensator (for a diffusion with drift $\mu$, $\dd A_t=e^{-rt}(rX_t-\mu(X_t))\dd t$; for geometric Brownian motion with convenience yield $q$, $\dd A_t=e^{-rt}qX_t\dd t$), the value \eqref{eq:Vintro} equals $\varphi(m)-X_0$, where $\varphi(m)=\sup_\tau\E[A_\tau+me^{-r\tau}]$ is Whittle's retirement value \cite{Whittle80} of the project ``hold the asset, collect the flow $\dd A$, and retire with reward $m$''. Hence $m^\ast$ is the retirement reward at which holding is worth exactly the spot, the signal $K$ is the Gittins index process \cite{Gittins79,Karatzas84,ElKarouiKaratzas94} of this project, and immediate delivery is optimal if and only if the index is at most the spot (Theorem~\ref{thm:gittins}). This is more than an analogy: the bounds
\[
X_0-\frac{1-\E[e^{-r\tau^\ast}]}{\E[e^{-r\tau^\ast}]}\,(K_0-X_0)^+\ \le\ m^\ast\ \le\ X_0\wedge\E[X_T],
\]
with $\tau^\ast$ the optimal delivery time, say that the discount of the futures price below spot never exceeds the excess of the index over spot times the expected financing cost of the delivery delay (Theorem~\ref{thm:gittins}(v)).

The ratio \eqref{eq:ratio} is handled throughout by Dinkelbach's device \cite{Dinkelbach67}, which reduces a fractional program to a parametric one and whose Newton iteration, $m_{k+1}=\E[e^{-r\tau_k}X_{\tau_k}]/\E[e^{-r\tau_k}]$ with $\tau_k$ optimal at the price $m_k$, we show to contract at rate $1-\E[e^{-rT}]$ (Theorem~\ref{thm:newton}); for a one-week window with $r\le5\%$, one standard optimal stopping problem at $m_0=X_0$ gives an iterate whose error is at most $10^{-3}(X_0-m^\ast)$.

\begin{figure}[t]\centering
\begin{tikzpicture}[>=Latex, node distance=4mm and 6mm, every node/.style={font=\footnotesize},
  box/.style={draw, rounded corners=2pt, align=center, inner sep=3pt, minimum height=6mm},
  main/.style={box, double, double distance=0.8pt, fill=blue!8, draw=blue!50!black},
  aux/.style={box, fill=black!4},
  lit/.style={box, dashed, fill=orange!12, draw=orange!60!black}]
\node[aux] (A1) {Assumption \ref{ass:standing}};
\node[aux, right=of A1] (P1) {Prop.\ \ref{prop:dinkelbach}\\Dinkelbach};
\node[lit, right=of P1] (BEK) {Lemma \ref{lem:bek}\\Bank--El Karoui};
\node[main, below=of P1] (T1) {Thm.\ \ref{thm:signal}\\signal $K$, $V'$, fixed point};
\node[aux, right=of T1] (P2) {Prop.\ \ref{prop:optimal}\\all optimal times};
\node[main, left=of T1] (T3) {Thm.\ \ref{thm:newton}\\Newton iteration};
\node[main, below=of T1] (T2) {Thm.\ \ref{thm:gittins}\\Gittins index, bounds};
\node[aux, left=of T2] (A2) {Assumption \ref{ass:semimart}};
\node[aux, below=of T2] (L3) {Lemma \ref{lem:monotone-m}\\index $=$ inverse boundary};
\node[aux, right=of L3] (P3) {Prop.\ \ref{prop:nearT}\\region near $T$};
\node[main, below=of L3] (T4) {Thm.\ \ref{thm:gbm}\\GBM};
\node[aux, left=of T4] (L4) {Lemma \ref{lem:brownian}\\Brownian problem, $C$};
\node[main, below=of T4] (T5) {Thm.\ \ref{thm:parabolic}\\$\sqrt{s}$ asymptotic};
\node[aux, left=of T5] (C1) {Cor.\ \ref{cor:Tc}\\$T_c$ near $q=r$};
\node[main, right=of T4] (T6) {Thm.\ \ref{thm:gs}\\Gibson--Schwartz};
\draw[->] (A1) -- (P1); \draw[->] (P1) -- (T1); \draw[->] (BEK) -- (T1);
\draw[->] (T1) -- (P2); \draw[->] (T1) -- (T3); \draw[->] (P1) -- (T3);
\draw[->] (T1) -- (T2); \draw[->] (A2) -- (T2);
\draw[->] (T2) -- (L3); \draw[->] (L3) -- (T4); \draw[->] (P3) -- (T4);
\draw[->] (L4) -- (T5); \draw[->] (T4) -- (T5); \draw[->] (T5) -- (C1);
\draw[->] (L3) -- (T6); \draw[->] (P3) -- (T6);
\end{tikzpicture}
\caption{Dependence of the results. Double borders: main theorems; single solid borders: auxiliary results; dashed borders: results quoted from the literature.}\label{fig:map}
\end{figure}
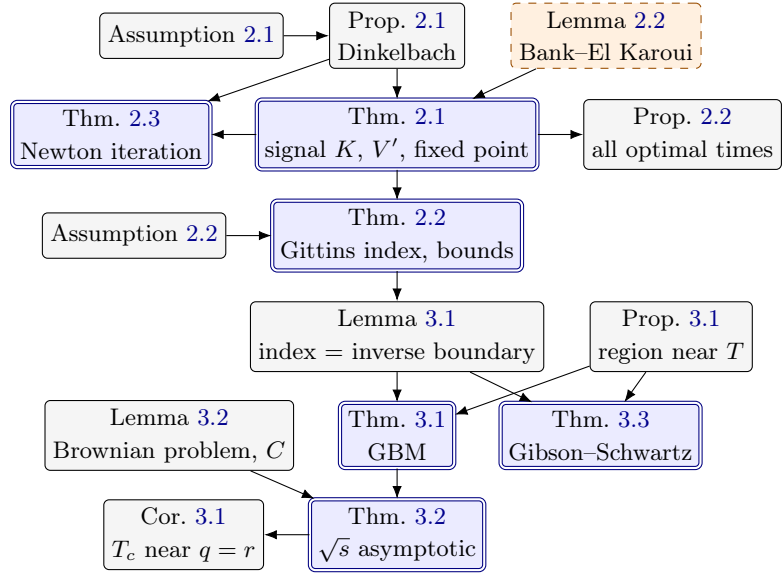

The second half of the paper applies this to two models. For geometric Brownian motion (Subsection~\ref{sec:gbm}) the futures price is a constant fraction $c^\ast$ of spot, and $c^\ast=1$ if and only if $q\le0$, or $0<q<r$ and the window is no longer than a critical horizon $T_c(r,q,\sigma)$. For $q>0$, let $B(s)$ denote the delivery boundary normalized to price $1$: with $s=T-t>0$ remaining and fixed delivery price $m>0$, it is optimal to deliver when $X_t\le mB(s)$ and to continue otherwise (Theorem~\ref{thm:gbm}). This normalized boundary approaches $r/q$ as $s\downarrow0$ and has the square-root asymptotic $B(s)=\frac rq(1-C\sigma\sqrt s)+o(\sqrt s)$, with a universal constant $C=0.6388\ldots$ characterized as the unique root of an explicit equation and as the value of a Brownian ratio problem (Theorem~\ref{thm:parabolic}); consequently $T_c\sim\big((1-q/r)/(C\sigma)\big)^2$ as $q\uparrow r$ (Corollary~\ref{cor:Tc}). For the Gibson--Schwartz model with stochastic convenience yield (Subsection~\ref{sec:gs}) the delivery region is monotone in the yield, and the price equals spot if and only if the initial yield is below a critical level that tends to $r$ as the window shrinks. Section~\ref{sec:numerics} reports numerical results, with convergence checks. Figure~\ref{fig:map} shows how the results depend on one another.

\textbf{Related work.} The obligation to exercise by expiry also appears in the FX ``open forward'' contract. Hok and Tse \cite{HokTse24} solve its exercise problem for a given fixed exchange rate under geometric Brownian motion, allowing negative interest rates. They characterize the exercise boundary and its near-expiry behaviour. A change of numeraire identifies their stopping problem with that of Subsection~\ref{sec:gbm} at a given delivery price (Remark~\ref{rem:hoktse}). Thus the boundary structure and square-root asymptotic have precedents in their work, the latter through the American-option analysis of Lamberton and Villeneuve \cite{LambertonVilleneuve03}. Our derivation identifies the coefficient as the value of a Brownian ratio problem and yields the asymptotic behaviour of the critical delivery-window length as $q\uparrow r$. The futures problem also requires the endogenous price $m^\ast$, selected by the zero-value condition \eqref{eq:zero}.

On the optimal stopping side, Bank and Baumgarten \cite{BankBaumgarten10} construct universal stopping signals for perpetual problems of regular one-dimensional diffusions with a parameter entering the reward linearly. They express the signal through transforms of level-passage times, give conditions for its monotonicity, and relate it to Gittins indices and stochastic representation. Here the Bank--El Karoui representation applies to an excess process that vanishes at the fixed expiry and gives a common delivery rule for all prices, including in non-Markovian models. The contract's zero-value condition then selects the price, and Dinkelbach's iteration computes it with the contraction bound of Theorem~\ref{thm:newton}.

The paper is organized as follows. Section~\ref{sec:theory} develops the general theory, from the ratio formulation to the universal delivery signal, its Gittins interpretation and the price iteration. Section~\ref{sec:models} studies the two models, and Section~\ref{sec:numerics} presents the numerical results. Section~\ref{sec:conclusion} discusses the remaining questions, and Appendix~\ref{app:num} describes the numerical methods.

\section{General theory}\label{sec:theory}

\subsection{Setting and Dinkelbach's reduction}\label{sec:setting}

\begin{assumption}\label{ass:standing}
\begin{enumerate}[label=(\roman*),leftmargin=2.2em]
\item $(\Omega,\F,\mathbb F=(\F_t)_{t\in[0,T]},\P)$ satisfies the usual conditions, $\F_0$ is trivial, and every $\mathbb F$-martingale has a continuous version (for instance, $\mathbb F$ is the augmented filtration of a Brownian motion). In particular every stopping time is either a.s.\ equal to $0$ or a.s.\ positive. $\P$ is a pricing measure and $T\in(0,\infty)$ is the length of the delivery window.
\item The interest rate $r=(r_t)_{t\in[0,T]}$ is progressively measurable with $r_t>0$ for all $t$, a.s., and $\int_0^Tr_t\dd t<\infty$ a.s. We write $D_t:=\exp(-\int_0^tr_s\dd s)$ and $d_T:=\E[D_T]\in(0,1)$.
\item The underlying $X=(X_t)_{t\in[0,T]}$ is continuous and adapted with $\E[\sup_{t\le T}|X_t|]<\infty$.
\end{enumerate}
\end{assumption}
$\cT[a,b]$ is the set of stopping times with values in $[a,b]$ and $\cT(a,b]$ the set of those with values in $(a,b]$ a.s. For $m\in\R$ set
\begin{equation}\label{eq:V}
V(m):=\sup_{\tau\in\cT[0,T]}\E\big[D_\tau(m-X_\tau)\big],\qquad
m^\ast:=\inf_{\tau\in\cT[0,T]}\frac{\E[D_\tau X_\tau]}{\E[D_\tau]} .
\end{equation}
Since $r>0$, $D$ is decreasing and $D_\tau\ge D_T$ for every $\tau$, so that $\E[D_\tau]\in[d_T,1]$. The following adapts Lindensj\"o's zero-value characterization \cite[Definition~3.1 and Proposition~3.2]{Lindensjo16} to Assumption~\ref{ass:standing}; the argument is supplied in Proposition~\ref{prop:dinkelbach} below.

\begin{lemma}[Lindensj\"o's zero-value characterization]\label{lem:lindensjo}
Under Assumption~\ref{ass:standing}, $m^\ast$ is the end-of-month futures price: it is the unique price at which the contract, with the short choosing the delivery time optimally, has zero value at the start of the window.
\end{lemma}

The following elementary proposition, which contains Lindensj\"o's argument, is Dinkelbach's reduction \cite{Dinkelbach67} of the fractional program \eqref{eq:ratio} to the parametric problem $V$.

\begin{proposition}\label{prop:dinkelbach}
Under Assumption~\ref{ass:standing}, $V$ is finite, convex and nondecreasing, with
\begin{equation}\label{eq:lipschitz}
d_T\,(m'-m)\ \le\ V(m')-V(m)\ \le\ m'-m\qquad(m<m'),
\end{equation}
$V(m)\to\pm\infty$ as $m\to\pm\infty$, and $m^\ast$ is the unique zero of $V$. A stopping time attains the infimum defining $m^\ast$ in \eqref{eq:V} if and only if it is optimal for $V(m^\ast)$.
\end{proposition}
\begin{proof}
For fixed $\tau$ the map $m\mapsto\E[D_\tau(m-X_\tau)]$ is affine with slope $\E[D_\tau]\in[d_T,1]$. A supremum of such maps is convex and satisfies the upper bound in \eqref{eq:lipschitz}; for the lower bound, given $\eps>0$ pick $\tau_\eps$ with $\E[D_{\tau_\eps}(m-X_{\tau_\eps})]\ge V(m)-\eps$ and note $V(m')\ge\E[D_{\tau_\eps}(m'-X_{\tau_\eps})]\ge V(m)-\eps+d_T(m'-m)$. Finiteness and the limits follow from \eqref{eq:lipschitz} and $|V(0)|\le\E\sup|X|$. Next, $V(m)\le0$ if and only if $\E[D_\tau(m-X_\tau)]\le0$ for every $\tau$, i.e.\ if and only if $m\le\E[D_\tau X_\tau]/\E[D_\tau]$ for every $\tau$, i.e.\ if and only if $m\le m^\ast$. So $\{V\le0\}=(-\infty,m^\ast]$ and, $V$ being continuous and strictly increasing, $m^\ast$ is its unique zero. Finally $\E[D_\tau(m^\ast-X_\tau)]=\E[D_\tau]\big(m^\ast-\E[D_\tau X_\tau]/\E[D_\tau]\big)\le0=V(m^\ast)$, with equality if and only if $\tau$ attains the infimum.
\end{proof}

\emph{Terminology.} Taking $\tau=0$ and $\tau=T$ in \eqref{eq:V} gives
\begin{equation}\label{eq:trivialbounds}
m^\ast\ \le\ X_0\wedge\frac{\E[D_TX_T]}{\E[D_T]} ,
\end{equation}
the two right-hand sides being the prices of the contract when delivery is forced at the start, respectively at the end, of the window (for a deterministic rate the second is the standard futures price $\E[X_T]$, and $m^\ast\le\E[X_T]$ is \cite[Prop.~3.2]{Lindensjo16} with $\tau=T$). The \emph{discount} $X_0-m^\ast\ge0$ of the futures price below spot measures what the short gains, relative to immediate delivery, from being free to delay; the difference $\E[D_TX_T]/\E[D_T]-m^\ast\ge0$ measures what she gains relative to forced delivery at $T$. Throughout, ``the futures price equals spot'' means $m^\ast=X_0$, equivalently (Proposition~\ref{prop:dinkelbach}) that immediate delivery is optimal; it does not mean that the delivery flexibility is worthless against the terminal benchmark, since then $\E[D_TX_T]/\E[D_T]-m^\ast$ is the carry, which may still be positive. Conversely, when the discount is positive it may consist largely of carry, as happens in Subsection~\ref{sec:gs} and Section~\ref{sec:numerics} for $q$ well above $r$.

\subsection{The universal delivery signal}\label{sec:signal}

Define the \emph{excess process}
\begin{equation}\label{eq:Z}
Z_t:=D_tX_t-\E\big[D_TX_T\,\big|\,\F_t\big],\qquad t\in[0,T],
\end{equation}
which is continuous (Assumption~\ref{ass:standing}(i)), of class (D), and satisfies $Z_T=0$. The tool is the representation theorem of Bank and El Karoui, which we recall in the form needed. An optional process $Y$ of class (D) is \emph{lower semicontinuous in expectation} if $\E[Y_S]\le\liminf_n\E[Y_{S_n}]$ for every monotone sequence of stopping times $S_n\to S$. Following \cite[Lemma~4.1]{BankElKaroui04}, an optional process $L$ with values in $[-\infty,\infty)$ is \emph{upper right-continuous} if
\[
L_t=\lim_{\eps\downarrow0}\sup_{s\in[t,(t+\eps)\wedge T)}L_s\qquad\text{for all }t\in[0,T)\text{ a.s.}
\]
The right envelope includes the current time; equivalently, $L$ is upper semicontinuous from the right.

\begin{lemma}[{Bank and El Karoui \cite[Theorems~1 and 3]{BankElKaroui04}; see also \cite[Section~2]{BankFollmer03}}]\label{lem:bek}
Let $f:\Omega\times[0,T]\times\R\to\R$ be such that $f(\omega,t,\cdot)$ is continuous and strictly decreasing from $+\infty$ to $-\infty$ for every $(\omega,t)$, $f(\cdot,\cdot,\ell)$ is progressively measurable, and $\E\int_0^T|f(t,\ell)|\dd t<\infty$ for every $\ell\in\R$. Let $Y$ be an optional process of class (D), lower semicontinuous in expectation, with $Y_T=0$. Then there is an optional process $L:\Omega\times[0,T)\to[-\infty,+\infty)$ such that
\begin{equation}\label{eq:BEK}
Y_S=\E\Big[\int_S^Tf\Big(t,\sup_{v\in[S,t)}L_v\Big)\dd t\;\Big|\;\F_S\Big]\qquad\text{for all }S\in\cT[0,T],
\end{equation}
with $f(t,\sup_{[S,t)}L)\1_{[S,T)}(t)\in L^1(\P\otimes\dd t)$. Among the upper right-continuous optional processes the solution of \eqref{eq:BEK} is unique up to indistinguishability, and this \emph{canonical} solution is given, on $\{S<T\}$, by
\begin{equation}\label{eq:Lformula}
L_S=\essinf_{\tau\in\cT(S,T]}\ell^{S,\tau},
\end{equation}
where $\ell^{S,\tau}$ is the unique $\F_S$-measurable random variable with
\[
\E[Y_S-Y_\tau\,|\,\F_S]=\E\Big[\int_S^\tau f(t,\ell^{S,\tau})\dd t\,\Big|\,\F_S\Big].
\]
The supremum in \eqref{eq:BEK} may equally be taken over $v\in[S,t]$: the running supremum of $L$ is nondecreasing in $t$ and can differ from its left limit only at countably many $t$, a $\dd t$-null set. The upper right-continuous regularization is justified by \cite[Lemma~4.1]{BankElKaroui04}.
\end{lemma}

In what follows $L$ always denotes the canonical solution, and we put
\begin{equation}\label{eq:Kdef}
K:=-L,\qquad\text{so that }K_t\le\liminf_{\substack{s\downarrow t\\s>t}}K_s\ \text{ for all }t\in[0,T)\text{ a.s.};
\end{equation}
by the d\'ebut theorem the first entrance time of the optional process $K$ into $(-\infty,m]$ is a stopping time for every $m$, and by the lower semicontinuity from the right in \eqref{eq:Kdef} this entrance time is attained whenever it lies before $T$: $K\le m$ there.

\begin{theorem}\label{thm:signal}
Let Assumption~\ref{ass:standing} hold.
\begin{enumerate}[label=(\roman*),leftmargin=2.2em]
\item Let $K$ be defined by \eqref{eq:Kdef} with $L$ the canonical solution of \eqref{eq:BEK} for $Y:=Z$ and $f(\omega,t,\ell):=-r_t(\omega)D_t(\omega)\ell$. Then $K$ is optional with values in $(-\infty,+\infty]$, satisfies \eqref{eq:Kdef}, and for every $S\in\cT[0,T]$
\begin{equation}\label{eq:rep}
Z_S=\E\Big[\int_S^Tr_tD_t\inf_{v\in[S,t)}K_v\dd t\;\Big|\;\F_S\Big],
\end{equation}
the integrand being in $L^1(\P\otimes\dd t)$; on $\{S<T\}$ it is given by
\begin{equation}\label{eq:Kformula}
K_S=\esssup_{\tau\in\cT(S,T]}\ \frac{X_S-\E\big[\tfrac{D_\tau}{D_S}X_\tau\,\big|\,\F_S\big]}{1-\E\big[\tfrac{D_\tau}{D_S}\,\big|\,\F_S\big]} .
\end{equation}
\item Put $I_t:=\inf_{v\in[0,t)}K_v$ for $t\in(0,T]$ and, for $m\in\R$, $\tau^m:=\inf\{t\in[0,T):K_t\le m\}\wedge T$, a stopping time with $K_{\tau^m}\le m$ on $\{\tau^m<T\}$. Then $\tau^m$ is optimal in \eqref{eq:V} and
\begin{equation}\label{eq:Vformula}
V(m)=\E\big[D_T(m-X_T)\big]+\E\int_0^Tr_tD_t\,(m-I_t)^+\dd t .
\end{equation}
\item For every $m$, $\E[D_{\tau^m}]$ lies between the one-sided derivatives of $V$ at $m$; in particular $V'(m)=\E[D_{\tau^m}]$ at every point of differentiability of $V$, i.e.\ for all but countably many $m$.
\item The end-of-month price is the unique solution of
\begin{equation}\label{eq:fixed}
\E\big[D_T(X_T-m^\ast)\big]=\E\int_0^Tr_tD_t\,(m^\ast-I_t)^+\dd t ,
\end{equation}
and $\tau^{m^\ast}$ is an optimal delivery time: deliver the first time the signal $K$ falls to the futures price, and at $T$ if it never does.
\end{enumerate}
\end{theorem}

\begin{proof}
(i) The function $f$ satisfies the requirements of Lemma~\ref{lem:bek}: it is strictly decreasing in $\ell$ because $r_tD_t>0$, and $\E\int_0^T|f(t,\ell)|\dd t=|\ell|\,\E[1-D_T]$ is finite. The process $Z$ is of class (D) and continuous; if $(S_n)$ is a monotone sequence of stopping times converging to $S$ then $Z_{S_n}\to Z_S$ a.s.\ and in $L^1$, so $Z$ is lower semicontinuous in expectation; and $Z_T=0$. Let $L$ be the canonical solution of \eqref{eq:BEK} and $K:=-L$, an optional process with values in $(-\infty,+\infty]$ satisfying \eqref{eq:Kdef}. Since $f(t,\sup_{[S,t)}L)=r_tD_t\inf_{[S,t)}K$, \eqref{eq:rep} is \eqref{eq:BEK}, and the integrability is part of the lemma. For the formula, the identity
\[
\E[Z_S-Z_\tau|\F_S]=\E[\int_S^\tau f(t,\ell)\dd t|\F_S]=-\ell\,\E[D_S-D_\tau|\F_S]
\]
gives
\[
\ell^{S,\tau}=-\frac{\E[Z_S-Z_\tau\,|\,\F_S]}{\E[D_S-D_\tau\,|\,\F_S]},
\]
By the tower property, the numerator equals $D_SX_S-\E[D_\tau X_\tau|\F_S]$. Thus $K_S=-L_S=\esssup_\tau(-\ell^{S,\tau})$, giving \eqref{eq:Kformula} after dividing numerator and denominator by $D_S$.

(ii) That $\tau^m$ is a stopping time with $K_{\tau^m}\le m$ on $\{\tau^m<T\}$ was noted after \eqref{eq:Kdef}. Fix $\tau\in\cT[0,T]$. Using $m(D_\tau-D_T)=\int_\tau^Tr_tD_tm\dd t$, $\E[Z_\tau]=\E[D_\tau X_\tau]-\E[D_TX_T]$ and \eqref{eq:rep} at $S=\tau$,
\begin{equation}\label{eq:gain}
\E\big[D_\tau(m-X_\tau)\big]=\E\big[D_T(m-X_T)\big]+\E\int_\tau^Tr_tD_t\Big(m-\inf_{v\in[\tau,t)}K_v\Big)\dd t .
\end{equation}
Pathwise, $\inf_{[\tau,t)}K\ge\inf_{[0,t)}K=I_t$ for $t>\tau$, so the integrand in \eqref{eq:gain} is at most $(m-I_t)^+$ and
\begin{equation}\label{eq:gainineq}
\int_\tau^Tr_tD_t\Big(m-\inf_{[\tau,t)}K\Big)\dd t\ \le\ \int_0^Tr_tD_t(m-I_t)^+\dd t .
\end{equation}
We claim that equality holds for $\tau=\tau^m$. If $t\le\tau^m$ then $K_v>m$ for all $v<t$ (as $v<\tau^m$), hence $I_t\ge m$ and $(m-I_t)^+=0$. If $t>\tau^m$ then, by the definition of $\tau^m$ as an infimum, there is $v\in[\tau^m,t)$ with $K_v\le m$; therefore $\inf_{[\tau^m,t)}K\le m\le\inf_{[0,\tau^m)}K$ and $\inf_{[\tau^m,t)}K=I_t=\min\big(\inf_{[0,\tau^m)}K,\inf_{[\tau^m,t)}K\big)$. On $\{\tau^m=T\}$ both sides vanish. Taking expectations in \eqref{eq:gain} proves the optimality of $\tau^m$ and \eqref{eq:Vformula}.

(iii) Pointwise, the map $m\mapsto(m-I_t)^+$ is convex with left derivative $\1_{\{I_t<m\}}$ and right derivative $\1_{\{I_t\le m\}}$. Its difference quotients are bounded by $1$. Dominated convergence with respect to the finite measure $r_tD_t\,\P(\dd\omega)\dd t$, whose total mass is $1-d_T$, therefore permits differentiation under $\E\int_0^Tr_tD_t(\cdot)\dd t$. Hence, from \eqref{eq:Vformula},
\[
\partial^-V(m)=\E[D_T]+\E\int_0^Tr_tD_t\1_{\{I_t<m\}}\dd t,\qquad
\partial^+V(m)=\E[D_T]+\E\int_0^Tr_tD_t\1_{\{I_t\le m\}}\dd t .
\]
Now $\{I_t<m\}\subset\{\tau^m<t\}\subset\{I_t\le m\}$ for $t\in(0,T]$: if $I_t<m$ there is $v<t$ with $K_v<m$, so $\tau^m\le v<t$; and if $\tau^m<t$ there is $v\in[\tau^m,t)$ with $K_v\le m$, so $I_t\le m$. Since $\E[D_T]+\E\int_0^Tr_tD_t\1_{\{\tau^m<t\}}\dd t=\E[D_T+\int_{\tau^m}^Tr_tD_t\dd t]=\E[D_{\tau^m}]$, we get $\partial^-V(m)\le\E[D_{\tau^m}]\le\partial^+V(m)$. A convex function is differentiable except at countably many points.

(iv) is (ii) and Proposition~\ref{prop:dinkelbach}: $V(m)=0$ rearranges to \eqref{eq:fixed}, whose left side is decreasing and whose right side is nondecreasing in $m$.
\end{proof}

The proof of (ii) gives the set of all optimal delivery times, in the form of \cite{BankFollmer03}.

\begin{proposition}\label{prop:optimal}
Under Assumption~\ref{ass:standing}, a stopping time $\tau$ is optimal for $V(m)$ if and only if, a.s., $K_v\ge m$ for all $v\in[0,\tau)$ and $K_\tau\le m$ on $\{\tau<T\}$. Equivalently: $\tau^m\le\tau\le\bar\tau^m:=\inf\{t:K_t<m\}\wedge T$ and $K_\tau=\inf_{v\in[0,\tau]}K_v$ on $\{\tau<T\}$.
\end{proposition}
\begin{proof}
By \eqref{eq:gain}, \eqref{eq:gainineq} and \eqref{eq:Vformula}, $\tau$ is optimal if and only if the expectations of the two sides of \eqref{eq:gainineq} agree; since the left integrand is at most the right one on $(\tau,T]$ and the right one is nonnegative on $[0,\tau]$, and since $r_tD_t>0$, this holds if and only if, a.s., (a) $(m-I_t)^+=0$ for a.e.\ $t\in(0,\tau]$ and (b) $m-\inf_{[\tau,t)}K=(m-I_t)^+$ for a.e.\ $t\in(\tau,T]$. The function $t\mapsto I_t$ is nonincreasing and left-continuous, so (a) is equivalent to $I_t\ge m$ for all $t\in(0,\tau]$, i.e.\ to $K_v\ge m$ for all $v\in[0,\tau)$. Given (a), $I_t=\min(\inf_{[0,\tau)}K,\inf_{[\tau,t)}K)$ equals $\inf_{[\tau,t)}K$ whenever the latter is at most $m$, and then (b) holds at $t$; whereas if $\inf_{[\tau,t)}K>m$ the left side of (b) is negative and the right side is zero. So, given (a), (b) is equivalent to $\inf_{[\tau,t)}K\le m$ for a.e., hence (by monotonicity in $t$) for all $t\in(\tau,T]$ on $\{\tau<T\}$, i.e.\ to $\lim_{t\downarrow\tau}\inf_{[\tau,t)}K=\min\big(K_\tau,\liminf_{s\downarrow\tau}K_s\big)\le m$, which by \eqref{eq:Kdef} is $K_\tau\le m$. The second formulation is a rewriting: $K\ge m$ on $[0,\tau)$ means $\tau\le\bar\tau^m$; and given $\tau\le\bar\tau^m$, the condition $K_\tau\le m$ on $\{\tau<T\}$ holds if and only if $\tau^m\le\tau$ and $K_\tau=\inf_{[0,\tau]}K$ there (if $K_\tau\le m$ then $\tau^m\le\tau$ by definition and $K_\tau\le m\le\inf_{[0,\tau)}K$; conversely, if $\tau^m\le\tau$ then $\inf_{[0,\tau]}K\le K_{\tau^m}\le m$).
\end{proof}

\begin{remark}\label{rem:signal}
(a) For $S=T$ formula \eqref{eq:Kformula} is void; $K_T$ never enters \eqref{eq:rep} or the stopping rule. (b) If $r$ is deterministic, \eqref{eq:fixed} reads $e^{-\int_0^Tr}(F^{\mathrm{std}}-m^\ast)=\E\int_0^Tr_te^{-\int_0^tr}(m^\ast-I_t)^+\dd t$ with $F^{\mathrm{std}}=\E X_T$ the standard futures price, and $m^\ast=F^{\mathrm{std}}$ if and only if $K\ge m^\ast$ on $[0,T)$ a.s. (c) The only properties of $K$ used beyond \eqref{eq:rep} are its optionality and the lower semicontinuity from the right in \eqref{eq:Kdef}, which make the crossing times $\tau^m$ stopping times at which $K\le m$ before $T$; the canonical version of Lemma~\ref{lem:bek} is singled out to have a definite process with these properties.
\end{remark}
\subsection{The Gittins index and computation of the price}\label{sec:gittins}

\begin{assumption}\label{ass:semimart}
$DX$ is a continuous semimartingale with canonical decomposition
\begin{equation}\label{eq:decomp}
D_tX_t=X_0+M_t-A_t,\qquad t\in[0,T],
\end{equation}
where $M$ is a uniformly integrable martingale and $A$ is a continuous adapted process of integrable total variation with $A_0=0$.
\end{assumption}
For a diffusion $\dd X=\mu(X)\dd t+\sigma(X)\dd W$ with constant $r$, $\dd A_t=e^{-rt}\big(rX_t-\mu(X_t)\big)\dd t$: $A$ represents the cumulative discounted holding reward, which may include non-cash benefits. Financing is accounted for separately by the discount factor. For geometric Brownian motion with net convenience yield $q$, $\dd A_t=e^{-rt}qX_t\dd t$; with stochastic convenience yield $\delta_t$, this becomes $\dd A_t=e^{-rt}\delta_tX_t\dd t$.

Consider the project ``hold the asset and collect $\dd A$'' with the option to \emph{retire} at a stopping time $\tau$ for a reward $m$, and its retirement value and Gittins index in the forward-induction form,
\begin{equation}\label{eq:whittle}
\varphi(m):=\sup_{\tau\in\cT[0,T]}\E\big[A_\tau+mD_\tau\big],\qquad
G_0:=\sup_{\tau\in\cT(0,T]}\frac{\E[A_\tau]}{\E[1-D_\tau]} .
\end{equation}
These are the objects of Gittins \cite{Gittins79} and Whittle \cite{Whittle80} for a bandit with discount $D$ and reward process $A$; see \cite{Karatzas84} for diffusions and \cite{ElKarouiKaratzas94,BankFollmer03} for the general continuous-time theory. The ratio defining $G_0$ is of the same kind as the ratio \eqref{eq:ratio} defining the price; the denominators differ: $\E[D_\tau]$ is the expected discount factor, while $\E[1-D_\tau]=\E\int_0^\tau r_tD_t\dd t$ is the expected discounted time, so that $m^\ast$ is the smallest discounted invoice per unit of discount factor and $G_0$ the largest reward per unit of discounted time, both over the same stopping times. Since $\E[A_\tau]=X_0-\E[D_\tau X_\tau]$ by \eqref{eq:decomp}, the two ratios are slopes of lines to the same set of points $(\E[D_\tau],\E[D_\tau X_\tau])$, measured from the origin and from the point $(1,X_0)$ of immediate delivery respectively; the theorem below says exactly how they are related, and Remark~\ref{rem:geometry} draws the picture.

\begin{theorem}\label{thm:gittins}
Let Assumptions~\ref{ass:standing} and \ref{ass:semimart} hold and let $K$ be the signal of Theorem~\ref{thm:signal}.
\begin{enumerate}[label=(\roman*),leftmargin=2.2em]
\item $V(m)=\varphi(m)-X_0$ for every $m\in\R$.
\item $K_S=\esssup_{\tau\in\cT(S,T]}\E[A_\tau-A_S\,|\,\F_S]\big/\E[D_S-D_\tau\,|\,\F_S]$ on $\{S<T\}$; in particular $K_0=G_0=\inf\{m:\varphi(m)=m\}$, and $\varphi(m)=m$ for $m\ge G_0$, $\varphi(m)>m$ for $m<G_0$.
\item $m^\ast$ is the unique solution of $\varphi(m^\ast)=X_0$. Immediate delivery is optimal, equivalently $m^\ast=X_0$, if and only if $K_0\le X_0$; otherwise $m^\ast<X_0<K_0$.
\item If $X_0<K_0<\infty$ then, with $\tau^u$ as in Theorem~\ref{thm:signal},
\begin{equation}\label{eq:identity}
\int_{m^\ast}^{K_0}\E[D_{\tau^u}]\dd u=K_0-X_0 .
\end{equation}
\item With $\tau^\ast:=\tau^{m^\ast}$ the optimal delivery time,
\begin{equation}\label{eq:bounds}
X_0-\Big(\frac1{\E[D_{\tau^\ast}]}-1\Big)(K_0-X_0)^+\ \le\ m^\ast\ \le\ X_0\wedge\frac{\E[D_TX_T]}{\E[D_T]},
\qquad\frac1{\E[D_{\tau^\ast}]}-1\le\frac1{d_T}-1 .
\end{equation}
\end{enumerate}
\end{theorem}
\begin{proof}
(i) By \eqref{eq:decomp} and optional sampling, $\E[D_\tau(m-X_\tau)]=m\E[D_\tau]-X_0+\E[A_\tau]$; take suprema.

(ii) By \eqref{eq:decomp}, $D_SX_S-\E[D_\tau X_\tau|\F_S]=\E[A_\tau-A_S|\F_S]$, which turns \eqref{eq:Kformula} (multiplied by $D_S$) into the stated formula. At $S=0$, since $\F_0$ is trivial, the essential supremum is a supremum of numbers and every $\tau\in\cT[0,T]$ is either $0$ or in $\cT(0,T]$; so $K_0=G_0$, using $\E[A_\tau]=X_0-\E[D_\tau X_\tau]$. Next, $\varphi(m)-m=\sup_{\tau\in\cT[0,T]}\E[A_\tau-m(1-D_\tau)]\ge0$ (take $\tau=0$), and $\varphi(m)-m\le0$ if and only if $\E[A_\tau]\le m\E[1-D_\tau]$ for all $\tau\in\cT(0,T]$ (the stopping time $0$ contributes $0$), i.e., since $\E[1-D_\tau]>0$ for such $\tau$, if and only if $m\ge G_0$. The function $\varphi(m)-m$ is nonincreasing because its slopes $\E[D_\tau]-1$ are $\le0$.

(iii) By (i) and Proposition~\ref{prop:dinkelbach}. Immediate delivery is optimal for $V(m^\ast)$ if and only if $V(m^\ast)=m^\ast-X_0$, i.e.\ $m^\ast=X_0$, i.e.\ $\varphi(X_0)=X_0$, i.e.\ $X_0\ge G_0=K_0$ by (ii). If $m^\ast<X_0$ then $\varphi(m^\ast)=X_0>m^\ast$, so $m^\ast<K_0$ by (ii); and if $K_0<\infty$ then $X_0=\varphi(m^\ast)<\varphi(K_0)=K_0$ because $\varphi$ is strictly increasing by \eqref{eq:lipschitz}, while for $K_0=\infty$ there is nothing to prove.

(iv) At the price $K_0$ we have $\tau^{K_0}=0$, so $V(K_0)=K_0-X_0$ by Theorem~\ref{thm:signal}(ii). Since $V(m^\ast)=0$, $m^\ast\le K_0$ (as $K_0-X_0>0=V(m^\ast)$ and $V$ is increasing), and $V$ is absolutely continuous with $V'=\E[D_{\tau^u}]$ a.e.\ by Theorem~\ref{thm:signal}(iii) and \eqref{eq:lipschitz}, we get $K_0-X_0=V(K_0)-V(m^\ast)=\int_{m^\ast}^{K_0}\E[D_{\tau^u}]\dd u$.

(v) The upper bounds are \eqref{eq:trivialbounds}. If $K_0\le X_0$ then $m^\ast=X_0$ by (iii). If $K_0=\infty$ the lower bound is void. Otherwise $u\mapsto\tau^u$ is nonincreasing, so $u\mapsto\E[D_{\tau^u}]$ is nondecreasing and \eqref{eq:identity} gives $(K_0-m^\ast)\E[D_{\tau^\ast}]\le K_0-X_0$, which rearranges to the lower bound; $\E[D_{\tau^\ast}]\ge d_T$.
\end{proof}

\begin{remark}[two ratios]\label{rem:geometry}
The signal \eqref{eq:Kformula} and the price \eqref{eq:ratio} are suprema and infima of different ratios over the same stopping times. Writing $d_\tau:=\E[D_\tau]$ and $a_\tau:=\E[D_\tau X_\tau]$, the point $(d_\tau,a_\tau)$ ranges over an ``attainable set'' in the plane containing $(1,X_0)$ (the point $\tau=0$); $m^\ast$ is the smallest slope of a line from the origin to the set, and $K_0$ the largest slope of a line from $(1,X_0)$ to a point of the set other than $(1,X_0)$ itself. Theorem~\ref{thm:gittins}(iii) says that there are exactly two configurations: either every such line has slope at most $X_0$ and then $m^\ast=X_0$, or some line is steeper than $X_0$ and then $m^\ast<X_0<K_0$, with \eqref{eq:identity} as the exact link when $K_0<\infty$. In the first configuration $K_0$ can be anything at most $X_0$: for geometric Brownian motion with $q=0$, $DX$ is a martingale, $A\equiv0$, $K\equiv0$ and $m^\ast=X_0$.
\end{remark}

\begin{remark}[Lindensj\"o's explicit cases]\label{rem:lindensjocases}
If $A$ is nonincreasing (a non-positive convenience yield, e.g.\ $\dd A\le0$ when $DX$ is a submartingale) then $\E[A_\tau]\le0$ for all $\tau$, hence $K\le0$ by (ii), and $m^\ast=X_0$ whenever $X_0\ge0$: this is \cite[Lemma 3.3]{Lindensjo16}. If $\dd A_t=D_tc\dd t$ with constants $c>0$ and $r$ (a continuous coupon) then $\E[A_\tau]/\E[1-D_\tau]=c/r$ for every $\tau$, so $K\equiv c/r$: deliver at once if $c\le rX_0$ and otherwise at $T$, in which case $m^\ast=X_0e^{rT}-\frac cr(e^{rT}-1)$ is the forward price; the lower bound in \eqref{eq:bounds} is then attained. This is the continuous-coupon analogue of the discrete-dividend case in \cite[Example~2]{Lindensjo16}.
\end{remark}

\begin{remark}[infinite index]\label{rem:infinite}
$K$ may take the value $+\infty$: by (ii) this happens at $S$ when the expected reward over short horizons is of larger order than the expected discounted time. The price $m^\ast$ is of course finite.
\end{remark}

\paragraph{Dinkelbach's Newton iteration.}
Theorem~\ref{thm:signal} characterizes $m^\ast$ through the running infimum of the index. For computation, the following iteration requires only standard optimal stopping problems, one per step, and converges fast on short windows.

\begin{theorem}\label{thm:newton}
Let Assumption~\ref{ass:standing} hold. Let $m_0\ge m^\ast$ (for instance $m_0=X_0$) and define recursively, with $\tau_k$ an optimal stopping time for $V(m_k)$,
\begin{equation}\label{eq:newton}
m_{k+1}:=\frac{\E[D_{\tau_k}X_{\tau_k}]}{\E[D_{\tau_k}]} .
\end{equation}
Then $m^\ast\le m_{k+1}\le m_k$ for all $k$ and
\begin{equation}\label{eq:rate}
0\le m_{k+1}-m^\ast\le(1-d_T)(m_k-m^\ast),\qquad\text{so}\qquad m_k-m^\ast\le(1-d_T)^k(m_0-m^\ast).
\end{equation}
If $V$ is differentiable at $m^\ast$, then either the iteration reaches $m^\ast$ in finitely many steps, or it converges superlinearly in the sense that
\[
\lim_{k\to\infty}\frac{m_{k+1}-m^\ast}{m_k-m^\ast}=0.
\]
\end{theorem}
\begin{proof}
By Theorem~\ref{thm:signal}(ii) optimal stopping times exist. From $V(m_k)=\E[D_{\tau_k}(m_k-X_{\tau_k})]=\E[D_{\tau_k}](m_k-m_{k+1})$ we get $m_{k+1}=m_k-V(m_k)/\E[D_{\tau_k}]$. As $m_k\ge m^\ast$, $V(m_k)\ge0$ and $m_{k+1}\le m_k$; and $m_{k+1}\ge m^\ast$ because it is a ratio in \eqref{eq:V}. By the lower bound in \eqref{eq:lipschitz}, $V(m_k)=V(m_k)-V(m^\ast)\ge d_T(m_k-m^\ast)$, hence $m_{k+1}-m^\ast\le(m_k-m^\ast)\big(1-d_T/\E[D_{\tau_k}]\big)\le(1-d_T)(m_k-m^\ast)$.

For the last claim, if $m_k=m^\ast$ for some $k$, then $V(m_k)=0$ and all subsequent iterates equal $m^\ast$. Otherwise $m_k>m^\ast$ for every $k$. Set $s_k:=\E[D_{\tau_k}]$. This is a subgradient of the convex function $V$ at $m_k$; differentiability at $m^\ast$ and $m_k\downarrow m^\ast$ imply $s_k\to V'(m^\ast)\ge d_T>0$. Since convexity gives $V(m_k)\ge V'(m^\ast)(m_k-m^\ast)$, the update yields
\[
0\le\frac{m_{k+1}-m^\ast}{m_k-m^\ast}
=1-\frac{V(m_k)}{s_k(m_k-m^\ast)}
\le1-\frac{V'(m^\ast)}{s_k}\longrightarrow0.
\]
\end{proof}

For a one-week window with $r\le5\%$, $1-d_T<10^{-3}$: the first iterate, which requires solving one optimal stopping problem at the price $X_0$, is already within $10^{-3}(X_0-m^\ast)$ of the price.
\section{Applications to specific models}\label{sec:models}

\subsection{Markovian underlyings: two general facts}\label{sec:markov}

In the rest of the paper $X$ is a function of a time-homogeneous diffusion. Let $O\subset\R^d$ be open, $b:O\to\R^d$, $\varsigma:O\to\R^{d\times d'}$ locally Lipschitz, and let $\Xi$ solve $\dd\Xi_t=b(\Xi_t)\dd t+\varsigma(\Xi_t)\dd W_t$ in $O$ without explosion. Let $\rho:O\to(0,\infty)$ be continuous (the short rate as a function of the state; $\rho\equiv r$ in Subsections~\ref{sec:gbm}--\ref{sec:gs}), $g\in C^2(O)$ (the delivery payoff, $g=m-x$ when $X=\Xi$ or its first coordinate), and
\begin{equation}\label{eq:markovV}
V(t,\xi):=\sup_{\tau\in\cT[t,T]}\E_{t,\xi}\Big[e^{-\int_t^\tau\rho(\Xi_s)\dd s}g(\Xi_\tau)\Big],\qquad
\cS:=\{V=g\},\quad\cC:=\{V>g\}\quad\text{in }[0,T)\times O,
\end{equation}
assuming $\E_{t,\xi}\sup_{s\in[t,T]}|g(\Xi_s)|<\infty$ for every $(t,\xi)$. We use three standard facts, valid whenever $V$ is continuous on $[0,T]\times O$ (which we verify in each model):
\begin{enumerate}[label=(F\arabic*),leftmargin=2.6em]
\item\label{F1} For $(t,\xi)\in\cC$ the first exit time $\tau_\cC:=\inf\{s\ge t:(s,\Xi_s)\notin\cC\}\wedge T$ is optimal, and $e^{-\int_t^{s}\rho(\Xi)}V(s,\Xi_s)$, $s\in[t,\tau_\cC]$, is a martingale \cite[Theorem~2.4, Remark~2.6 and Corollary~2.9]{PeskirShiryaev06}.
\item\label{F2} $V(\cdot,\xi)$ is nonincreasing (time-homogeneity), so $\cS_t:=\{\xi:(t,\xi)\in\cS\}$ is nondecreasing in $t$.
\item\label{F3} $\cS$ is closed, $\cC$ is open, and $V(T,\cdot)=g$.
\end{enumerate}
\ref{F1} and \ref{F3} only use that the reward process is continuous and of class (D) and that the value function is continuous.
Write $\cL:=b\cdot\nabla+\tfrac12\operatorname{tr}(\varsigma\varsigma^\top D^2)$ for the generator and
\begin{equation}\label{eq:f}
f:=\cL g-\rho g\in C(O)
\end{equation}
for the running reward of holding: by Dynkin's formula, $e^{-\int\rho}g(\Xi_\tau)-g(\xi)=\int_0^\tau e^{-\int\rho}f(\Xi_s)\dd s+$ a martingale, for bounded stopping times before the exit from a compact set.

\begin{proposition}[the delivery region near the end of the window]\label{prop:nearT}
Assume $V$ is continuous on $[0,T]\times O$.
\begin{enumerate}[label=(\roman*),leftmargin=2.2em]
\item If $f(\xi_0)>0$ then there is $\delta>0$ with $(t,\xi_0)\in\cC$ for all $t\in[T-\delta,T)$.
\item If $f<0$ on a neighbourhood of $\xi_0$ then there is $\delta>0$ with $(t,\xi_0)\in\cS$ for all $t\in[T-\delta,T)$. If $Q\subset O$ is compact and $f<0$ on a neighbourhood of $Q$, then $\delta$ can be chosen uniformly for $\xi_0\in Q$.
\end{enumerate}
\end{proposition}
\begin{proof}
(i) Let $B$ be a ball around $\xi_0$ with $\overline B\subset O$ and $f\ge c>0$ on $\overline B$, and let $\bar\rho:=\max_{\overline B}\rho$. For $h\in(0,T-t]$ let $\theta:=\inf\{s\ge t:\Xi_s\notin B\}\wedge(t+h)$. By Dynkin's formula (the coefficients and $g$ are bounded on $\overline B$),
$V(t,\xi_0)\ge\E_{t,\xi_0}[e^{-\int_t^\theta\rho}g(\Xi_\theta)]=g(\xi_0)+\E\int_t^\theta e^{-\int_t^s\rho}f(\Xi_s)\dd s\ge g(\xi_0)+ce^{-\bar\rho h}\E[\theta-t]>g(\xi_0)$, since $\theta>t$ a.s.

(ii) Let $U$ be an open ball of radius $\varrho$ around $\xi_0$ with $\overline U\subset O$ and $f\le-c<0$ on $\overline U$, let $\bar\rho:=\max_{\overline U}\rho$, and $u:=V-g\ge0$, which is continuous on $[0,T]\times O$ with $u(T,\cdot)=0$. Fix $t\in[0,T)$, $\tau\in\cT[t,T]$, and let $\theta:=\inf\{s\ge t:\Xi_s\notin U\}\wedge T$ and $\vartheta:=\tau\wedge\theta-t$. Dynkin's formula on $[t,\tau\wedge\theta]$ gives $\E[e^{-\int_t^{\tau\wedge\theta}\rho}g(\Xi_{\tau\wedge\theta})]=g(\xi_0)+\E\int_t^{\tau\wedge\theta}e^{-\int_t^s\rho}f(\Xi_s)\dd s\le g(\xi_0)-ce^{-\bar\rho(T-t)}\E[\vartheta]$. On $\{\theta<\tau\}$, by the strong Markov property at $\theta$ and the definition of $V$, $\E[e^{-\int_\theta^\tau\rho}g(\Xi_\tau)|\F_\theta]\le V(\theta,\Xi_\theta)=g(\Xi_\theta)+u(\theta,\Xi_\theta)$. Combining,
\begin{equation}\label{eq:nearTbound}
\E\big[e^{-\int_t^\tau\rho}g(\Xi_\tau)\big]\le g(\xi_0)-ce^{-\bar\rho(T-t)}\E[\vartheta]+\E\big[\1_{\{\theta<\tau\}}u(\theta,\Xi_\theta)\big].
\end{equation}
On $\{\theta<\tau\}$ we have $\theta<T$, $\Xi_\theta\in\partial U$ and $\sup_{s\in[t,t+\vartheta]}|\Xi_{s\wedge\theta}-\xi_0|=\varrho$. The coefficients being bounded on $\overline U$, the Burkholder--Davis--Gundy inequality applied to the stopped process gives $\E\sup_{s\in[t,t+\vartheta]}|\Xi_{s\wedge\theta}-\xi_0|^4\le C_1\E[\vartheta^2]$ with $C_1$ depending only on the coefficient bounds on $\overline U$ and on $T$; hence $\P(\theta<\tau)\le C_1\varrho^{-4}\E[\vartheta^2]\le C_1\varrho^{-4}(T-t)\E[\vartheta]$. Moreover $u(\theta,\Xi_\theta)\le\bar u(T-t)$ on $\{\theta<\tau\}$, where $\bar u(\delta):=\max\{u(s,\xi):s\in[T-\delta,T],\ \xi\in\partial U\}\to0$ as $\delta\downarrow0$, by uniform continuity of $u$ on the compact set $[0,T]\times\partial U$ and $u(T,\cdot)=0$. Therefore
\[
\E\big[e^{-\int_t^\tau\rho}g(\Xi_\tau)\big]-g(\xi_0)\le\E[\vartheta]\Big(-ce^{-\bar\rho(T-t)}+\bar u(T-t)\,C_1\varrho^{-4}(T-t)\Big)\le0
\]
as soon as $T-t\le\delta$, where $\delta>0$ is chosen so that $\bar u(\delta)C_1\delta/\varrho^4\le ce^{-\bar\rho\delta}$. Taking the supremum over $\tau$ gives $V(t,\xi_0)=g(\xi_0)$ for $t\in[T-\delta,T)$. If $f\le-c$ on a compact neighbourhood $N$ of a compact $Q$, the balls of a fixed radius $\varrho$ around the points of $Q$ lie in $N$ for $\varrho$ small, and $c$, $\bar\rho$, $C_1$ and $\bar u$ (defined with $N$ in place of $\partial U$) can be taken uniform over $\xi_0\in Q$.
\end{proof}

The second fact concerns the family of problems indexed by the delivery price $m$. Let $x$ be the underlying-price coordinate of $\xi$, put $g_m(\xi):=m-x$, and write $V(t,\xi;m)$ for the value in \eqref{eq:markovV} with payoff $g=g_m$. The corresponding regions are $\cS_m:=\{V(\cdot,\cdot;m)=g_m\}$ and $\cC_m:=\{V(\cdot,\cdot;m)>g_m\}$. Define the delivery-price threshold by
\begin{equation}\label{eq:markovK}
K(t,\xi):=\inf\{m\in\R:V(t,\xi;m)=m-x\}
=\inf\{m\in\R:(t,\xi)\in\cS_m\},
\end{equation}
for $(t,\xi)\in[0,T)\times O$, with $\inf\varnothing:=+\infty$. This defines a deterministic function of time and state taking values in $[-\infty,+\infty]$. The next lemma identifies it with the optional signal process $K_s$ of Theorem~\ref{thm:signal} when evaluated along the diffusion.

\begin{lemma}\label{lem:monotone-m}
Assume $V(\cdot,\cdot;m)$ is continuous on $[0,T]\times O$ for every $m$. Then $m\mapsto V(t,\xi;m)-(m-x)$ is nonincreasing and $\cS_m$ is nondecreasing in $m$. The function $K$ defined in \eqref{eq:markovK} is Borel and satisfies $\cS_m=\{K\le m\}$. For the problem started at $(t,\xi)$, the optional signal has the representation
\[
K_s=K(s,\Xi_s),\qquad s\in[t,T),
\]
up to indistinguishability.
\end{lemma}
\begin{proof}
For fixed $\tau$ the map $m\mapsto\E[e^{-\int\rho}(m-X_\tau)]-(m-x)$ has slope $\E[e^{-\int\rho}]-1\le0$; take the supremum. If $(t,\xi)\in\cS_{m_n}$ with $m_n\downarrow m$ then $V(t,\xi;m)=m-x$ by continuity of $V$ in $m$ (Proposition~\ref{prop:dinkelbach}), so $\{m:(t,\xi)\in\cS_m\}$ is a closed half-line $[K,\infty)$, and $K=\inf\{m\in\mathbb Q:V(t,\xi;m)=m-x\}$ is Borel since $V$ is continuous. Finally, by \eqref{eq:Kformula}, for every $m$ and every stopping time $S<T$ the event $\{K_S\le m\}$ is, up to a null set, $\{\E[D_\tau(m-X_\tau)|\F_S]\le D_S(m-X_S)\text{ for all }\tau\in\cT(S,T]\}$, the event that stopping at $S$ is optimal at the price $m$ for the problem started at $S$; by the strong Markov property that problem has value $V(S,\Xi_S;m)$ \cite[Chapter~I]{PeskirShiryaev06}, so $\{K_S\le m\}=\{(S,\Xi_S)\in\cS_m\}=\{K(S,\Xi_S)\le m\}$ a.s.\ for every rational $m$, whence $K_S=K(S,\Xi_S)$ a.s.\ for every stopping time $S<T$. Both processes being optional ($K(\cdot,\Xi_\cdot)$ is Borel in $(s,\omega)$ and adapted with continuous $\Xi$), they are indistinguishable by the optional section theorem.
\end{proof}
Thus $K(t,\xi)$ is the delivery-price threshold at each time and state. For a contract started at time $0$, the running infimum of this signal along the state path is
\[
I_t:=\inf_{0\le v<t}K(v,\Xi_v),\qquad 0<t\le T.
\]
The process $I$ enters the pricing equation \eqref{eq:fixed}. The futures price is the scalar $m^\ast$ uniquely determined by that equation, as stated in Theorem~\ref{thm:signal}(iv).

\subsection{Geometric Brownian motion}\label{sec:gbm}

Under the pricing measure, let $\dd X_t=(r-q)X_t\dd t+\sigma X_t\dd W_t$, $X_0=x>0$, with constants $r>0$, $q\in\R$ and $\sigma>0$. Here $q$ is the underlying asset\textquotesingle s \emph{net holding-yield rate}, not a payment from the futures contract. For a commodity, it is the net convenience yield: $qX_t$ represents the value per unit time of having the physical inventory available, less storage costs. This benefit need not be a cash payment; for a dividend-paying asset, $q$ instead plays the role of the dividend yield. Financing is accounted for separately by $r$, so $q<0$ represents a net holding cost. In this case, and also when $q=0$, immediate delivery is optimal and $m^\ast=X_0$, as Theorem~\ref{thm:gbm} below shows: at any positive delivery price $m$, the running gain from waiting is $qX_t-rm<0$. The discounted reward entering the holding project of Subsection~\ref{sec:gittins} is $\dd A_t=e^{-rt}qX_t\dd t$.

By scaling, $V(t,x;m)=m\,V(t,x/m;1)$ for $m>0$, so it suffices to study the price $1$. We write $s:=T-t$ for the time to the end of the window and $V(s,x)$ for the value with price $1$ and time-to-go $s$. Define the stopping region from the value function, and then its upper endpoint, by
\begin{equation}\label{eq:gbmregion}
\cS_s:=\{x>0:V(s,x)=1-x\}\quad(s\ge0),\qquad
B(s):=\sup\cS_s\quad(s>0).
\end{equation}
Theorem~\ref{thm:gbm}(i) below proves that $\cS_s$ is a nonempty interval, so $B(s)$ is its upper delivery boundary. Here $s$ is the remaining time: $\cS_s$ is the section of the stopping set of Subsection~\ref{sec:markov} at calendar time $T-s$. At expiry $V(0,x)=1-x$, hence $\cS_0=(0,\infty)$; $B$ is defined only for $s>0$, and $B(0+)$ denotes its limit as $s\downarrow0$.

Let $x^\ast(s)$ denote the unique zero of $V(s,\cdot)$, whose existence is verified below. The \emph{normalized futures price} is the futures price as a fraction of the initial spot price:
\begin{equation}\label{eq:cstar}
c^\ast=c^\ast(r,q,\sigma,T):=\frac{m^\ast}{X_0}=\frac{1}{x^\ast(T)}.
\end{equation}
The last equality follows from scaling and the zero-value condition: $V(T,X_0/m^\ast)=0$. Thus $c^\ast=1$ means that the futures price equals spot, while $1-c^\ast$ is the discount below spot as a fraction of $X_0$.
Let $\gamma>0$ be the positive root of $\tfrac{\sigma^2}2\gamma(\gamma+1)-(r-q)\gamma-r=0$ and $b_\infty:=\gamma/(1+\gamma)$ the perpetual American put boundary with strike $1$.

\begin{theorem}\label{thm:gbm}
\begin{enumerate}[label=(\roman*),leftmargin=2.2em]
\item $V$ is continuous on $[0,\infty)\times(0,\infty)$, convex and nonincreasing in $x$, and nondecreasing in $s$.

For every $s>0$, the stopping region $\cS_s$ is a nonempty interval. If $q\le0$, then $\cS_s=(0,\infty)$ and $B(s)=+\infty$. If $q>0$, then
\[
\cS_s=(0,B(s)],\qquad 0<B(s)<\infty,
\]
and $B$ is nonincreasing and upper semicontinuous on $(0,\infty)$.
\item If $q>0$: $B(s)\le r/q$ for all $s>0$, $B(0+):=\lim_{s\downarrow0}B(s)=r/q$, and $B(\infty):=\lim_{s\to\infty}B(s)=b_\infty<1$.
\item The normalized futures price $c^\ast$ defined in \eqref{eq:cstar} belongs to $(0,1]$, is nonincreasing in $T$, $q$ and $\sigma$, and satisfies $c^\ast\le\min(1,e^{(r-q)T})$. If $q\le0$, then $c^\ast=1$. If $q\ge r$, then $c^\ast<1$ for every $T>0$. In the remaining case $0<q<r$, the critical horizon satisfies
\[
T_c:=\sup\{s>0:B(s)\ge1\}\in(0,\infty),\qquad
c^\ast=1\quad\Longleftrightarrow\quad T\le T_c.
\]
\item For $q\ge0$, the signal is $K(t,x)=x/B(T-t)$ for $t<T$, with $x/(+\infty):=0$. For every $q\in\R$, an optimal delivery time is
\[
\tau^\ast=\inf\{0\le t<T:X_t\le c^\ast X_0\,B(T-t)\}\wedge T.
\]
In particular, $\tau^\ast=0$ when $q\le0$. The quantities $c^\ast$ and $\sigma^2T_c$ (when $0<q<r$) depend on $(r,q,\sigma,T)$ only through $(r/\sigma^2,q/\sigma^2,\sigma^2T)$.
\end{enumerate}
\end{theorem}
\begin{proof}
(i) Write $X^x_t=x\hat X_t$ with $\hat X$ the process started at $1$. Then $V(s,x)=\sup_{\tau\le s}\E[e^{-r\tau}(1-x\hat X_\tau)]$ is a supremum of affine nonincreasing functions of $x$, hence convex and nonincreasing, and $|V(s,x)-V(s,x')|\le|x-x'|\,\E\sup_{t\le s}\hat X_t$. In $s$ it is nondecreasing (more stopping times), and for $s<s'$, $0\le V(s',x)-V(s,x)\le\E\sup_{s\le t\le s'}|e^{-rt}(1-x\hat X_t)-e^{-rs}(1-x\hat X_s)|\to0$ as $s'\downarrow s$ or $s\uparrow s'$ by continuity of paths and dominated convergence. So $V$ is continuous. For $q>0$, $e^{-rt}\hat X_t$ is a supermartingale, hence $\E[e^{-r\tau}\hat X_\tau]\le1$ and each affine map $x\mapsto\E[e^{-r\tau}(1-x\hat X_\tau)]-(1-x)$ has slope $1-\E[e^{-r\tau}\hat X_\tau]\ge0$; so $V(s,\cdot)-(1-\cdot)$ is nondecreasing and $\cS_s$ is an interval $(0,B(s)]$ (closed in $(0,\infty)$ because $\cS$ is closed, and nonempty because $V(s,\cdot)\le V_\infty$ with $V_\infty$ the perpetual put value of (ii) below, which equals $1-x$ on $(0,b_\infty]$). $B$ is nonincreasing because $\cS_s$ shrinks as $s$ grows, and upper semicontinuous because $\cS$ is closed. If $q\le0$ then $e^{-rt}\hat X_t$ is a submartingale and $\E[e^{-r\tau}(1-x\hat X_\tau)]\le\E[e^{-r\tau}]-x\le1-x$, so $\cS_s=(0,\infty)$. If $q>0$ then for $x$ large $V(s,x)\ge\E[e^{-rs}(1-x\hat X_s)]=e^{-rs}-xe^{-qs}>1-x$, so $B(s)<\infty$.

(ii) Here $f(x)=(\cL-r)(1-x)=qx-r$. By Proposition~\ref{prop:nearT}(i), every $x>r/q$ lies in $\cC_s$ for small $s$, hence for all $s>0$ (as $\cS_s$ shrinks in $s$), which gives $B(s)\le r/q$. By Proposition~\ref{prop:nearT}(ii), every $x_0<r/q$ (where $f<0$ on a neighbourhood) lies in $\cS_s$ for small $s$, so $B(s)\ge x_0$ for small $s$; hence $B(0+)=r/q$, the limit existing by monotonicity. For the limit at infinity let $V_\infty(x):=\sup_{\tau\in\cT[0,\infty]}\E[e^{-r\tau}(1-x\hat X_\tau)\1_{\{\tau<\infty\}}]$. Clearly $V(s,x)\le V_\infty(x)$, and for every $\tau\in\cT[0,\infty]$, $\E[e^{-r(\tau\wedge s)}(1-x\hat X_{\tau\wedge s})]\to\E[e^{-r\tau}(1-x\hat X_\tau)\1_{\{\tau<\infty\}}]$ as $s\to\infty$, because $e^{-rs}\to0$ and $\E[e^{-rs}\hat X_s]=e^{-qs}\to0$ ($q>0$); hence $V(s,x)\uparrow V_\infty(x)$. Moreover $V_\infty$ is the perpetual American put price with strike $1$: replacing $\tau$ by $\tau$ on $\{x\hat X_\tau\le1\}$ and $+\infty$ otherwise does not decrease the payoff, so $V_\infty(x)=\sup_\tau\E[e^{-r\tau}(1-x\hat X_\tau)^+]$, whose stopping region is $(0,b_\infty]$ \cite[Section~25.1]{PeskirShiryaev06}. Since $V(s,\cdot)\le V_\infty$, $\cS_s\supset(0,b_\infty]$, so $B(\infty)\ge b_\infty$; and if $x\in(b_\infty,B(\infty))$ then $V(s,x)=1-x$ for all $s$, whence $V_\infty(x)=1-x$ and $x\le b_\infty$, a contradiction. Finally $b_\infty<1$.

(iii) For each finite $s$, continuity and strict positivity of $\hat X$ give
\[
\ell_s:=\E\Big[\inf_{0\le t\le s}e^{-rt}\hat X_t\Big]>0.
\]
Every affine function defining $V(s,\cdot)$ has slope at most $-\ell_s$, so
$V(s,x')\le V(s,x)-\ell_s(x'-x)$ for $x'>x\ge0$, with $V(s,0)=1$ by continuous extension. Thus $V(s,x)\le1-\ell_s x$, and continuity gives a unique zero $x^\ast(s)\in[1,\infty)$, since $V(s,1)\ge0$. Scaling and Proposition~\ref{prop:dinkelbach} now give $m^\ast=X_0/x^\ast(T)>0$.

Monotonicity in $T$ is monotonicity of $V$ in $s$ (the zero $x^\ast(T)$ of the nonincreasing function $V(T,\cdot)$ moves right as $V$ increases); in $q$ it follows from the pathwise comparison $X^{q}_t=X^{0}_te^{-qt}$ (a larger $q$ lowers $X$, raises the payoff $1-X$, and hence lowers $c^\ast$); in $\sigma$ it follows from the convexity of the payoff $x\mapsto1-x$ and the monotonicity of American values with convex payoffs in the volatility \cite{ElKarouiJeanblancShreve98,Hobson98}. The bounds are \eqref{eq:trivialbounds}. Now $c^\ast=1$ if and only if $V(T,1)=0$ if and only if $1\in\cS_T$, i.e.\ $B(T)\ge1$. If $q\le0$, $B\equiv\infty$. If $q>r$, $c^\ast\le e^{(r-q)T}<1$. If $q=r$, then $\hat X$ is a martingale and for $a\in(0,1)$ and $\tau_a:=\inf\{t:\hat X_t\le1-a\}\wedge T$,
\[
V(T,1)\ \ge\ \E\big[e^{-r\tau_a}(1-\hat X_{\tau_a})\big]=r\,\E\int_0^{\tau_a}e^{-rt}(\hat X_t-1)\dd t
\]
(It\^o's formula for $e^{-rt}(1-\hat X_t)$ and optional sampling), and $\E\int_0^{\tau_a}(\hat X_t-1)\dd t=\E[\tau_a(\hat X_{\tau_a}-1)]=a\,\E[(T-\tau_a)^+]>0$ by integration by parts, $\E\int_0^{\tau_a}t\dd\hat X_t=0$, and $\E[(\hat X_T-1)\1_{\{\tau_a=T\}}]=-\E[(\hat X_{\tau_a}-1)\1_{\{\tau_a<T\}}]=a\P(\tau_a<T)$; while $|\E\int_0^{\tau_a}(1-e^{-rt})(\hat X_t-1)\dd t|\le rT^2\E\sup_{t\le T}|\hat X_t-1|$. Choose $a=\sqrt T$: by Brownian scaling $\P(\tau_a\le T/2)\to\P(\inf_{u\le1/2}(\sigma W_u)\le-1)>0$ as $T\downarrow0$, so $\E[(T-\tau_a)^+]\ge(T/2)\P(\tau_a\le T/2)\ge cT$ for small $T$, while the discounting error is $O(T^{5/2})$. Hence $V(T,1)>0$ for small $T$, and for all $T$ by monotonicity, i.e.\ $c^\ast<1$. If $0<q<r$: $B(0+)=r/q>1>b_\infty=B(\infty)$ by (ii), so $T_c\in(0,\infty)$; and $B(T)\ge1$ if and only if $T\le T_c$, because $B$ is nonincreasing and upper semicontinuous (if $B(s_n)\ge1$ with $s_n\uparrow T_c$ then $B(T_c)\ge1$).

(iv) Suppose first that $q>0$. For a positive delivery price $m$, scaling gives the stopping region $(0,mB(s)]$. At a nonpositive price $m$, the running reward $qx-rm$ is strictly positive, so immediate delivery is not optimal before expiry. Lemma~\ref{lem:monotone-m} therefore gives $K(t,x)=\inf\{m>0:x\le mB(T-t)\}=x/B(T-t)$. If $q=0$, the compensator $A$ vanishes, so $K\equiv0$, consistently with $B\equiv+\infty$. For $q\ge0$ the delivery rule follows from Theorem~\ref{thm:signal}; for $q<0$ it gives $\tau^\ast=0$, which is optimal by (i). The scaling follows from the time change $t\mapsto\sigma^2t$, under which the problem becomes the same problem with parameters $(r/\sigma^2,q/\sigma^2,1,\sigma^2T)$.
\end{proof}

\begin{remark}[relation to the FX open forward]\label{rem:hoktse}
For a fixed positive delivery price $m$, the following change of numeraire relates our stopping problem to the FX open forward. It also provides the measure change used in Proposition~\ref{prop:lindensjo} below.

Under the measure $\P^X$ with density $e^{-(r-q)T}X_T/X_0$ and with $Y:=1/X$, Bayes' rule gives $\E[e^{-r\tau}(m-X_\tau)]=X_0m\,\E^X[e^{-q\tau}(Y_\tau-1/m)]$, and $Y$ is a geometric Brownian motion with drift $q-r$ under $\P^X$. So the short's obligation problem is, under the foreign numeraire, the long's problem for an FX open forward with domestic rate $q$, foreign rate $r$ and fixed rate $1/m$, which is the problem solved by Hok and Tse \cite{HokTse24} for a given fixed rate, including negative rates; their exercise boundary corresponds to $B$ and their degenerate cases to $q\le0$ and $r\le0$. What is specific to the futures contract is the fixed point $m^\ast$ and its consequences (iii)--(iv).
\end{remark}

Theorem~\ref{thm:gbm} determines the futures price when the end-of-month window begins. The next proposition inserts this price into Lindensj\"o's contract, showing how the end-of-month option affects the quoted futures price and delivery decisions during the preceding timing-option period.

\begin{proposition}[Lindensj\"o's contract for geometric Brownian motion]\label{prop:lindensjo}
Consider Lindensj\"o's contract \cite[Section 4]{Lindensjo16}: no option on $[0,T_t)$, the timing option on $[T_t,T_e]$ and the end-of-month option on $[T_e,T]$, with $c^\ast=c^\ast(r,q,\sigma,T-T_e)$ the fraction of Theorem~\ref{thm:gbm}. Then the futures price process is
\[
F_t=X_t\min\big\{1,\ c^\ast e^{(r-q)(T_e-t)}\big\}\ \ (r\ge q),\qquad F_t=c^\ast X_te^{(r-q)(T_e-t)}\ \ (r<q),\qquad t\in[T_t,T_e],
\]
and $F_t=\E_t[F_{T_t}]$ for $t<T_t$, i.e.\ $F_t=X_te^{(r-q)(T_t-t)}\min\{1,c^\ast e^{(r-q)(T_e-T_t)}\}$ for $r\ge q$ and $F_t=c^\ast X_te^{(r-q)(T_e-t)}$ for $r<q$.
\end{proposition}
\begin{proof}
By \cite[Prop.~4.6]{Lindensjo16}, on $[T_t,T_e]$, $F_t=\inf_{\tau\in\cT[t,T_e]}\E_t[Y_\tau]$ with $Y=X$ on $[T_t,T_e)$ and $Y_{T_e}=c^\ast X_{T_e}$, i.e.\ $F_t=\inf_\tau\E_t[X_\tau(1-(1-c^\ast)\1_{\{\tau=T_e\}})]$. Under $\P^X$ of Remark~\ref{rem:hoktse} this is $X_t\inf_\tau\E^X_t[e^{(r-q)(\tau-t)}(1-(1-c^\ast)\1_{\{\tau=T_e\}})]$, the infimum of the expectation of a deterministic function $\psi(\tau)$ of the stopping time alone; since $\E^X_t[\psi(\tau)]\ge\inf_{u\in[t,T_e]}\psi(u)$ and deterministic times attain it, $F_t=X_t\min\{\inf_{u\in[t,T_e)}e^{(r-q)(u-t)},c^\ast e^{(r-q)(T_e-t)}\}$, which is the claim. For $t<T_t$ use \cite[Prop.~4.9]{Lindensjo16}.
\end{proof}
With $r>q$, the short delivers at the very start of the delivery month unless the end-of-month discount outweighs the carry over the month, $1/c^\ast>e^{(r-q)(T_e-T_t)}$; with $r<q$ the short waits for the end-of-month window in all cases.

\subsubsection{The delivery boundary near the end of the window}\label{sec:parabolic}

The boundary $B$ of Theorem~\ref{thm:gbm} approaches $r/q$ at square-root rate as the remaining time tends to zero, with a universal constant that we characterize as the root of an explicit equation. The constant comes from a Brownian stopping problem, which we solve first.

Let $W$ be a standard Brownian motion in its augmented filtration and consider
\begin{equation}\label{eq:wdef}
w(h,z):=\sup_{\tau\in\cT[0,h]}\E\int_0^\tau(z+W_u)\dd u,\qquad h\ge0,\ z\in\R .
\end{equation}
Define the auxiliary function
\[
g(z):=e^{-z^2/2}\int_0^\infty t^3e^{-zt-t^2/2}\dd t.
\]

\begin{lemma}[the Brownian problem]\label{lem:brownian}
The equation $g(-c)+c\,g'(-c)=0$ has a unique root $C>0$, with $C=0.638833\ldots$. Define
\begin{equation}\label{eq:fdef}
f(\zeta):=\left(\zeta-\frac{g(\zeta)}{g'(-C)}\right)\1_{\{\zeta>-C\}} .
\end{equation}
For $h>0$, the value and stopping region are
\begin{equation}\label{eq:wform}
w(h,z)=h^{3/2}f\big(z/\sqrt h\big),\qquad
\{z:w(h,z)=0\}=(-\infty,-C\sqrt h].
\end{equation}
An optimal stopping time is
\[
\tau^\ast:=\inf\{u\in[0,h):z+W_u\le-C\sqrt{h-u}\}\wedge h.
\]
Moreover, the constant $C$ has the ratio representation
\begin{equation}\label{eq:Cratio}
C=\sup_{\tau\in\cT(0,1]}\frac{\E\int_0^\tau W_u\dd u}{\E[\tau]} .
\end{equation}
\end{lemma}
\begin{proof}
\emph{The auxiliary function and the root.} Write
\[
I_k(z):=\int_0^\infty t^ke^{-zt-t^2/2}\dd t,\qquad g(z)=e^{-z^2/2}I_3(z).
\]
Differentiation under the integral gives $g\in C^\infty(\R)$. Since $\partial_t e^{-zt-t^2/2}=-(z+t)e^{-zt-t^2/2}$, integration by parts and differentiation give
\[
I_{k+1}+zI_k=kI_{k-1}\quad(k\ge1),\qquad I_k'=-I_{k+1}.
\]
Consequently,
\begin{equation}\label{eq:gderiv}
\begin{aligned}
g'&=-e^{-z^2/2}(I_4+zI_3)=-3e^{-z^2/2}I_2<0,\\
g''&=6e^{-z^2/2}I_1>0,\qquad g'''=-6e^{-z^2/2}I_0<0.
\end{aligned}
\end{equation}
The same recurrence yields the ODE:
\[
g''+zg'-3g=-3e^{-z^2/2}(I_3+zI_2-2I_1)=0.
\]
Clearly $g>0$ and $g\to0$ at $+\infty$.

To identify the root, put $\phi(c):=g(-c)+cg'(-c)$. Then
\[
\phi(0)=g(0)>0,\qquad \phi'(c)=-cg''(-c)<0\quad(c>0).
\]
By Laplace's method,
\[
g(-c)\sim\sqrt{2\pi}\,c^3,\qquad
g'(-c)\sim-3\sqrt{2\pi}\,c^2\qquad(c\to\infty).
\]
Thus $\phi(c)\to-\infty$, and the root $C$ exists and is unique.

\emph{The candidate profile.} Put $A:=-1/g'(-C)>0$, so $f(\zeta)=\zeta+Ag(\zeta)$ for $\zeta>-C$ and $f=0$ otherwise. By construction,
\[
f(-C+)=-C-\frac{g(-C)}{g'(-C)}=0,\qquad
f'(-C+)=1+Ag'(-C)=0.
\]
Hence $f\in C^1(\R)$ and $f$ is $C^2$ off $-C$. On $(-C,\infty)$, $f''=Ag''>0$, so $f'$ increases from $f'(-C)=0$ to $\lim_{\zeta\to\infty}f'=1+A\lim g'=1$. Thus $0\le f'\le1$ on $\R$ and $f>0$ on $(-C,\infty)$. The ODE for $g$ and the definition of $C$ give
\[
\begin{aligned}
f''(-C+)&=3Ag(-C)+CAg'(-C)=3C-C=2C,\\
f''+\zeta f'-3f+2\zeta&=A(g''+\zeta g'-3g)=0.
\end{aligned}
\]
Since $g'''<0$, the second derivative decreases on $(-C,\infty)$, so $0<f''\le2C$ there. Finally, $g\to0$ implies $f(\zeta)-\zeta\to0$ as $\zeta\to\infty$.

\emph{Verification.} Let $\tilde w(h,z):=h^{3/2}f(z/\sqrt h)$ for $h>0$ and $\tilde w(0,\cdot):=0$. This function is continuous on $[0,\infty)\times\R$, nonnegative, $C^1$ on $(0,\infty)\times\R$ by smooth fit, and $C^{1,2}$ off the curve $z=-C\sqrt h$. With $\zeta=z/\sqrt h$, its derivatives are
\[
\tilde w_z=hf'(\zeta),\qquad \tilde w_{zz}=\sqrt h\,f''(\zeta).
\]
Both are bounded on $(0,h_0]\times\R$, and on this strip
\[
|\partial_h\tilde w|=\sqrt h\,\big|\tfrac32f(\zeta)-\tfrac12\zeta f'(\zeta)\big|\le c(1+|z|).
\]
A direct computation gives
\begin{equation}\label{eq:HJBw}
-\partial_h\tilde w+\tfrac12\tilde w_{zz}+z=\tfrac{\sqrt h}2\big(f''+\zeta f'-3f+2\zeta\big)(\zeta)=\begin{cases}0,&z>-C\sqrt h,\\ z<0,&z<-C\sqrt h.\end{cases}
\end{equation}
Fix $(h,z)$, $h>0$, and $\tau\in\cT[0,h]$. Since $\tilde w$ is $C^1$ and $C^{1,2}$ off a continuous curve of bounded variation across which $\tilde w_z$ is continuous, the change-of-variable formula of \cite{Peskir05} applies to $\tilde w(h-u,z+W_u)$ with no local-time term, and the stochastic integral is a true martingale because $\tilde w_z$ is bounded; so
\begin{align*}
\E\big[\tilde w(h-\tau,z+W_\tau)\big]
&=\tilde w(h,z)+\E\int_0^\tau\big(-\partial_h\tilde w+\tfrac12\tilde w_{zz}\big)(h-u,z+W_u)\dd u\\
&\le\tilde w(h,z)-\E\int_0^\tau(z+W_u)\dd u
\end{align*}
by \eqref{eq:HJBw}. Since $\tilde w\ge0$, it follows that
\[
\E\int_0^\tau(z+W_u)\dd u
\le\tilde w(h,z)-\E[\tilde w(h-\tau,z+W_\tau)]
\le\tilde w(h,z).
\]
If $z\le-C\sqrt h$, the preceding bound gives $w(h,z)=\tilde w(h,z)=0$, attained by $\tau^\ast=0$. If $z>-C\sqrt h$, the path stays in $\{z'>-C\sqrt{h'}\}$ before $\tau^\ast$, where \eqref{eq:HJBw} is an equality, and $\tilde w(h-\tau^\ast,z+W_{\tau^\ast})=0$. Indeed, on $\{\tau^\ast<h\}$ the point lies on the curve, while $\tilde w(0,\cdot)=0$ handles $\tau^\ast=h$. Thus equality holds and $w=\tilde w$, with $\tau^\ast$ optimal, proving \eqref{eq:wform}.

\emph{The ratio characterization.} We have
\[
w(1,z)=\sup_\tau\left(z\E[\tau]+\E\int_0^\tau W_u\dd u\right).
\]
Since $\F_0$ is trivial, every stopping time other than $0$ is a.s.\ positive. Therefore $w(1,z)=0$ if and only if
\[
z\le-\frac{\E\int_0^\tau W_u\dd u}{\E[\tau]}
\qquad\text{for all }\tau\in\cT(0,1].
\]
Comparing with $\{w(1,\cdot)=0\}=(-\infty,-C]$ gives \eqref{eq:Cratio}.
\end{proof}

\begin{theorem}[square-root boundary asymptotic]\label{thm:parabolic}
Let $q>0$ and let $B(s)$ be the delivery boundary for geometric Brownian motion in Theorem~\ref{thm:gbm}, for delivery price $1$ and remaining time $s$. Put $\bar x:=r/q$, and let $C$ be the constant from the auxiliary Brownian stopping problem in Lemma~\ref{lem:brownian}. Then
\begin{equation}\label{eq:parabolic}
B(s)=\bar x\big(1-C\sigma\sqrt s\big)+o(\sqrt s)\qquad(s\downarrow0),
\end{equation}
and the $o(\sqrt s)$ is uniform for $q$ in compact subsets of $(0,\infty)$. Equivalently, the delivery index satisfies $K(T-s,x)=\frac qr\,x\,\big(1+C\sigma\sqrt s+o(\sqrt s)\big)$.
\end{theorem}
\begin{proof}
\emph{Scaling.} Let $u(s,x):=V(s,x)-(1-x)=\sup_{\tau\in\cT[0,s]}\E_x\int_0^\tau e^{-rt}(qX_t-r)\dd t$ (Theorem~\ref{thm:gittins}(i)). Put
\[
x=\bar x(1+\sigma\sqrt s\,z),\qquad
Z^s_v:=\frac{X_{sv}-\bar x}{\bar x\sigma\sqrt s},\qquad v\in[0,1].
\]
Throughout this proof, $Z^s$ denotes this centered and rescaled GBM price, with state space $I_s:=(-1/(\sigma\sqrt s),\infty)$ and initial state $z\in I_s$. We have $Z^s_0=z$ and $qX_{sv}-r=q\bar x\sigma\sqrt s\,Z^s_v$. The time change $t=sv$ maps $\cT[0,s]$ onto $\cT[0,1]$ in the rescaled filtration $(\F_{sv})_{v\in[0,1]}$, hence
\[
u\big(s,\bar x(1+\sigma\sqrt s\,z)\big)=q\bar x\sigma s^{3/2}\,u_s(0,z),\qquad
u_s(v,z):=\sup_{\tau\in\cT[v,1]}\E\Big[\int_v^\tau e^{-rsu}Z^s_u\dd u\,\Big|\,Z^s_v=z\Big],
\]
where $u_s(v,z)$, $(v,z)\in[0,1]\times I_s$, is the value of the same problem started at time $v$ from $z$, with the discount factor $e^{-rsu}$ counted from time $0$ throughout; thus $u_s\ge0$, $u_s(1,\cdot)=0$, and by the strong Markov property $u_s(\theta,Z^s_\theta)$ dominates the conditional expected reward of any continuation beyond a stopping time $\theta$. Since $\cS_s=(0,B(s)]$, the set $\{z\in I_s:u_s(0,z)=0\}$ is $(-1/(\sigma\sqrt s),\beta_s]$ with $\beta_s:=(B(s)-\bar x)/(\bar x\sigma\sqrt s)$, and \eqref{eq:parabolic} is the statement $\beta_s\to-C$.

\emph{Step 1: rescaling the GBM dynamics.} The process $\tilde W_v:=s^{-1/2}W_{sv}$ is Brownian in the rescaled filtration. Applying the time and space changes to $\dd X_t=(r-q)X_t\dd t+\sigma X_t\dd W_t$ gives $\dd Z^s_v=b_s(Z^s_v)\dd v+\sqrt{a_s(Z^s_v)}\,\dd\tilde W_v$, where
\[
a_s(z)=(1+\sigma\sqrt s\,z)^2,\qquad b_s(z)=\sigma^{-1}(r-q)\sqrt s\,(1+\sigma\sqrt s\,z).
\]
These specific coefficients converge locally uniformly to $1$ and $0$, respectively, as $s\downarrow0$. The following estimate quantifies the resulting approximation of $Z^s$ by $z+\tilde W$.
Write $\alpha_v:=\sigma\sqrt s\tilde W_v+(r-q-\sigma^2/2)sv$. Then
\[
Z^s_v-(z+\tilde W_v)
=\frac{e^{\alpha_v}-1-\alpha_v}{\sigma\sqrt s}
+\frac{r-q-\sigma^2/2}{\sigma}\sqrt s\,v
+z(e^{\alpha_v}-1).
\]
Also, $|e^{-rsv}Z^s_v-Z^s_v|\le rs|Z^s_v|$. The bounds
\[
|e^\alpha-1-\alpha|\le\tfrac12\alpha^2e^{|\alpha|},\qquad
|e^\alpha-1|\le|\alpha|e^{|\alpha|},
\]
together with the exponential moments of $\sup_{v\le1}|\tilde W_v|$, give a constant $C_0=C_0(r,q,\sigma)$ such that, for $s\le1$,
\begin{equation}\label{eq:Zerr}
\begin{aligned}
\E\sup_{v\le1}\big|Z^s_v-(z+\tilde W_v)\big|&\le C_0(1+|z|)\sqrt s,\\
\E\sup_{v\le1}\big|e^{-rsv}Z^s_v-(z+\tilde W_v)\big|&\le C_0(1+|z|)\sqrt s.
\end{aligned}
\end{equation}

\emph{Step 2: $\limsup\beta_s\le-C$.} Fix $z>-C$ and let $\tau'$ be the optimal time for $w(1,z)$ of Lemma~\ref{lem:brownian}, a stopping time of the filtration of $\tilde W$. By \eqref{eq:Zerr},
\[
\begin{aligned}
u_s(0,z)&\ge\E\int_0^{\tau'}e^{-rsv}Z^s_v\dd v\\
&\ge w(1,z)-C_0(1+|z|)\sqrt s.
\end{aligned}
\]
Since $w(1,z)=f(z)>0$, this is positive for sufficiently small $s$. Hence $z>\beta_s$ for small $s$.

\emph{Step 3: $\liminf\beta_s\ge-C$.} Fix $\eta\in(0,1)$ and define
\[
U(v,z):=w(1-v,z+\eta),\qquad (v,z)\in[0,1]\times\R.
\]
The estimates in the proof of Lemma~\ref{lem:brownian} give $U\ge0$ and $U(1,\cdot)=0$. For $v<1$, the function is $C^1$ everywhere and $C^{1,2}$ off the curve $z=-C\sqrt{1-v}-\eta$, with
\[
0\le U_z\le1,\qquad
0\le U_{zz}\le F_2:=\sup_{\zeta>-C}f''(\zeta)\le2C,
\]
where the second-derivative bound applies off the curve. By \eqref{eq:HJBw},
\begin{equation}\label{eq:Ueq}
\begin{aligned}
\partial_vU+\tfrac12U_{zz}+(z+\eta)&=0
&&\text{on }\{z+\eta>-C\sqrt{1-v}\},\\
U&=0
&&\text{on }\{z+\eta\le-C\sqrt{1-v}\}.
\end{aligned}
\end{equation}
To define the comparison domain, put
\[
c_1:=\frac{\bar x-b_\infty}{\bar x\sigma}.
\]
The quadratic for $\gamma$ gives $(q-r)\gamma<r$, hence $b_\infty<\bar x$ and $0<c_1<1/\sigma$. Fix $M>0$, to be chosen below, and set
\[
D:=[0,1)\times(-c_1/\sqrt s,M).
\]
On $D$, $a_s\ge(b_\infty/\bar x)^2>0$. We compare $u_s$ with $U$ on this domain.

(a) \emph{Boundary.} At $z=-c_1/\sqrt s$, equivalently $x=b_\infty$, Theorem~\ref{thm:gbm}(ii) gives $u_s=0\le U$. At $v=1$, both functions vanish.

At the right boundary, let $Z^{s,M}$ denote the process started at $M$. Time-homogeneity, $e^{-rsu}\le1$, and \eqref{eq:Zerr} give
\begin{align*}
u_s(v,M)&\le\E\int_0^{1-v}(Z^{s,M}_u)^+\dd u\\
&\le\E\int_0^{1-v}(M+\tilde W_u)^+\dd u+(1-v)C_0(1+M)\sqrt s\\
&\le(1-v)\big[M+\varepsilon(M)+C_0(1+M)\sqrt s\big],
\end{align*}
where $\varepsilon(M):=\E(M+\tilde W_1)^-\le e^{-M^2/2}$. Waiting until the terminal time in the Brownian problem gives the lower bound
\[
U(v,M)=w(1-v,M+\eta)
\ge\E\int_0^{1-v}(M+\eta+\tilde W_u)\dd u
=(1-v)(M+\eta).
\]
First choose $M$ so that $\varepsilon(M)\le\eta/2$. Then choose $s_1(\eta,M)>0$ so that $C_0(1+M)\sqrt s\le\eta/2$ for $s\le s_1(\eta,M)$. The two bounds give $U\ge u_s$ on $\{z=M\}$.

(b) \emph{Supersolution.} Off the curve, set
\[
G_s:=\partial_vU+\tfrac12a_sU_{zz}+b_sU_z+e^{-rsv}z.
\]
Where $z+\eta<-C\sqrt{1-v}$, we have $U=0$ and $G_s=e^{-rsv}z<0$. Above the curve, \eqref{eq:Ueq} gives
\[
G_s=\tfrac12(a_s-1)U_{zz}+b_sU_z+(e^{-rsv}-1)z-\eta .
\]
We bound the three error terms separately. On $D$, $|z|\le\max(c_1/\sqrt s,M)$. For $s\le(\sigma M)^{-2}$, this gives
\[
\sigma\sqrt s|z|\le\sigma c_1+1,\qquad
|b_s|\le K_1\sqrt s,\qquad
K_1:=\sigma^{-1}|r-q|(2+\sigma c_1).
\]
The discounting term satisfies
\[
(e^{-rsv}-1)z\le rs|z|\1_{\{z<0\}}\le rc_1\sqrt s.
\]
For the second-derivative term, take $s<M^{-4}$ so that $\{z\ge s^{-1/4}\}\cap D$ is empty. On $\{z<-s^{-1/4}\}\cap D$, the inequality $z+\eta<-C\sqrt{1-v}$ holds for small $s$, and hence $U=0$ there. On the remaining set $\{|z|\le s^{-1/4}\}$,
\[
|a_s-1|\le2\sigma s^{1/4}+\sigma^2\sqrt s,\qquad
\tfrac12|a_s-1|U_{zz}\le K_2s^{1/4},\qquad
K_2:=(\sigma+\sigma^2/2)F_2.
\]
Combining these bounds, on $D$ and off the curve,
\[
G_s\le K_2s^{1/4}+K_1\sqrt s+rc_1\sqrt s-\eta<0\qquad\text{for }s\le s_0(\eta,M).
\]

(c) \emph{Comparison.} Fix $z\in(-c_1/\sqrt s,M)$ and $\tau\in\cT[0,1]$, and let
\[
\theta:=\inf\{v:Z^s_v\notin(-c_1/\sqrt s,M)\}\wedge1.
\]
Apply the change-of-variable formula of \cite{Peskir05} to $U(v,Z^s_v)$ on $[0,\tau\wedge\theta\wedge(1-\epsilon)]$. There is no local-time term, the derivatives are bounded on $\overline D$, and $Z^s$ spends no time on the curve since $a_s>0$ on $D$. Letting $\epsilon\downarrow0$ is justified by dominated convergence: $U$ is continuous and bounded on $\overline D$, with $U(1,\cdot)=0$. Part (b) therefore gives
\begin{align*}
\E\big[U(\tau\wedge\theta,Z^s_{\tau\wedge\theta})\big]
&=U(0,z)+\E\int_0^{\tau\wedge\theta}\big(\partial_vU+\tfrac12a_sU_{zz}+b_sU_z\big)\dd v\\
&\le U(0,z)-\E\int_0^{\tau\wedge\theta}e^{-rsv}Z^s_v\dd v.
\end{align*}
On $\{\theta<\tau\}$, the point $(\theta,Z^s_\theta)$ lies on the lateral boundary of $D$. The strong Markov property at $\theta$ and part (a) give
\begin{align*}
\E\bigg[\1_{\{\theta<\tau\}}\int_\theta^\tau e^{-rsv}Z^s_v\dd v\bigg]
&\le\E\big[\1_{\{\theta<\tau\}}u_s(\theta,Z^s_\theta)\big]\\
&\le\E\big[\1_{\{\theta<\tau\}}U(\theta,Z^s_\theta)\big].
\end{align*}
Adding the two inequalities and using $U\ge0$,
\[
\E\int_0^\tau e^{-rsv}Z^s_v\dd v\le U(0,z)-\E\big[U(\tau,Z^s_\tau)\1_{\{\tau\le\theta\}}\big]\le U(0,z)=f(z+\eta).
\]
Taking the supremum over $\tau$, $u_s(0,z)\le f(z+\eta)$, which vanishes for $z\le-C-\eta$. Hence $\beta_s\ge-C-\eta$ for $s\le s_0(\eta,M)\wedge s_1(\eta,M)$. Together with Step 2, $\beta_s\to-C$.

To make the uniformity explicit, fix a compact set $J\subset(0,\infty)$ of yields. The positive continuous functions $\bar x(q)$, $b_\infty(q)$ and $c_1(q)$ have positive minima and finite maxima on $J$. In particular, write $\underline c_1:=\min_{q\in J}c_1(q)>0$ and $\overline c_1:=\max_{q\in J}c_1(q)<\infty$. The exponential-moment bounds in Step 1 use $q$ only through bounded drift coefficients, so \eqref{eq:Zerr} holds with one constant $\overline C_0$ for all $q\in J$; likewise $K_1$ is bounded by one $\overline K_1$, while $K_2$ is independent of $q$. Step 2 is therefore uniform for each fixed $z>-C$. In Step 3, choose $M$ using $\eta$ alone, and then choose $s$ small enough that
\[
\overline C_0(1+M)\sqrt s\le\eta/2,\qquad
K_2s^{1/4}+(\overline K_1+r\overline c_1)\sqrt s<\eta,
\qquad \underline c_1/\sqrt s>C+\eta,
\]
together with the restrictions on $s$ in part (b), which are independent of $q$. These choices ensure boundary domination, the supersolution inequality, and inclusion of $z=-C-\eta$ in every comparison domain. Thus $\beta_s\to-C$ uniformly on $J$; since $\bar x$ is bounded on $J$, the remainder in \eqref{eq:parabolic} is uniform there. The statement about $K$ is Theorem~\ref{thm:gbm}(iv).
\end{proof}

\begin{corollary}[the critical horizon near $q=r$]\label{cor:Tc}
As $q\uparrow r$,
\[
T_c(r,q,\sigma)=\Big(\frac{1-q/r}{C\sigma}\Big)^2\big(1+o(1)\big)=\Big(\frac{r/q-1}{C\sigma}\Big)^2\big(1+o(1)\big).
\]
\end{corollary}
\begin{proof}
Fix $\eta\in(0,C)$. By Theorem~\ref{thm:parabolic} there is $s_0>0$ such that $\bar x(1-(C+\eta)\sigma\sqrt s)\le B(s)\le\bar x(1-(C-\eta)\sigma\sqrt s)$ for all $s\le s_0$ and $q\in[r/2,r]$. Since $\bar x=r/q\downarrow1$ as $q\uparrow r$, the upper bound gives $B(s_0)<1$ for $q$ close to $r$, so $T_c<s_0$ by monotonicity of $B$; then $B(s)\ge1$ requires $\sqrt s\le(1-q/r)/((C-\eta)\sigma)$ and holds whenever $\sqrt s\le(1-q/r)/((C+\eta)\sigma)$, i.e.\ $\big((1-q/r)/((C+\eta)\sigma)\big)^2\le T_c\le\big((1-q/r)/((C-\eta)\sigma)\big)^2$. Let $\eta\downarrow0$. The second form follows from $(r/q-1)/(1-q/r)=r/q\to1$.
\end{proof}

\begin{remark}\label{rem:parabolic}
(a) The Brownian limit suggests an extension to a one-dimensional diffusion $\dd X=\mu(X)\dd t+\nu(X)\dd W$. If the running reward $f(x)=rx-\mu(x)-rm$ in \eqref{eq:f} vanishes at $\bar x$, with $f'(\bar x)\ne0$ and $\nu(\bar x)\ne0$, the local scaling predicts $B(s)=\bar x-\operatorname{sgn}(f'(\bar x))\,C|\nu(\bar x)|\sqrt s+o(\sqrt s)$ for a delivery boundary emanating from $\bar x$. The proof above uses the explicit GBM solution and the stopping-region properties of Theorem~\ref{thm:gbm}. An extension would require suitable coefficient regularity, integrability estimates and one-sided boundary assumptions; we do not prove it here. (b) The limit problem \eqref{eq:wdef} is the one that governs the American put near expiry when the dividend yield exceeds the interest rate, in which case the exercise boundary also starts at the zero of the running reward; the square-root asymptotic in that setting was derived in \cite{EvansKuskeKeller02} and proved in \cite{LambertonVilleneuve03}. In the notation of \cite[Section~2.1, Lemma~1]{LambertonVilleneuve03}, $\phi(y)=w(1,y)$ and the stopping threshold is $y^\ast=-C$; \cite[Theorem~2]{LambertonVilleneuve03} therefore gives the same coefficient $C$ in the square-root asymptotic. Lemma~\ref{lem:brownian} supplies its explicit root characterization. (c) In the language of Subsection~\ref{sec:gittins}, Theorem~\ref{thm:parabolic} says that the Gittins index of holding exceeds its ``static'' value $\frac qr x$ by the factor $1+C\sigma\sqrt s$: the excess is the value of delivering adaptively, and $C$ is the Dinkelbach value \eqref{eq:Cratio} of the Brownian ratio ``reward collected per unit time''.
\end{remark}
\subsection{Stochastic convenience yield: the Gibson--Schwartz model}\label{sec:gs}

Let, under the pricing measure,
\begin{equation}\label{eq:gs}
\dd X_t=(r-\delta_t)X_t\dd t+\sigma X_t\dd W^1_t,\qquad \dd\delta_t=\kappa(\alpha-\delta_t)\dd t+\sigma_\delta\dd W^2_t,\qquad \dd\langle W^1,W^2\rangle_t=\rho\dd t,
\end{equation}
with constants $r,\sigma,\sigma_\delta,\kappa>0$, $\alpha\in\R$, $\rho\in(-1,1)$ \cite{GibsonSchwartz90,Schwartz97}. Here $\delta_t$ is the net convenience-yield rate: inventory benefits less storage costs, with financing accounted for separately by $r$. The process $A$ in the discounted-price decomposition \eqref{eq:decomp} is the cumulative discounted net holding reward:
\[
A_t=\int_0^t e^{-ru}\delta_uX_u\dd u,\qquad A_0=0.
\]
Since $\delta$ can have either sign, $A$ need not be increasing. Let $V(t,x,\delta;m)$ be the value \eqref{eq:markovV} with $\Xi=(X,\delta)$, $\rho\equiv r$, $g=m-x$. Auxiliary delivery prices $m$ range over $\R$, so the normalized price $y=m/x$ also ranges over $\R$.

\begin{theorem}\label{thm:gs}
\begin{enumerate}[label=(\roman*),leftmargin=2.2em]
\item $V$ is continuous and scales as $V(t,x,\delta;m)=x\,v(t,m/x,\delta)$, where we write $v(t,y,\delta):=V(t,1,\delta;y)$.
\item $v-(y-1)$ is nonincreasing in $y\in\R$ and nondecreasing in $\delta$. Consequently there are functions $d(t,y)\in[-\infty,\infty]$, nondecreasing in $y$ and in $t$, and $k(t,\delta)\in(-\infty,+\infty]$, nondecreasing in $\delta$ and nonincreasing in $t$, with
\[
\cS=\{(t,y,\delta):\delta\le d(t,y)\}=\{(t,y,\delta):y\ge k(t,\delta)\}\quad(t<T).
\]
\item $d(t,y)\le ry$ for all $t<T$, $y\in\R$, and $\lim_{t\uparrow T}d(t,y)=ry$.
\item $m^\ast=c^\ast(\delta_0)X_0$ with $c^\ast(\delta_0)=k^\ast(\delta_0)\in(0,1]$, where $k^\ast(\delta_0)$ is the zero of $v(0,\cdot,\delta_0)$; $c^\ast$ is nonincreasing in $\delta_0$; and $c^\ast(\delta_0)=1$ if and only if $\delta_0\le\delta_c(T):=d(0,1)$. The critical yield $\delta_c(T)\le r$ is nonincreasing in $T$ and $\delta_c(T)\to r$ as $T\downarrow0$.
\item The signal is $K(t,x,\delta)=x\,k(t,\delta)$, and the optimal delivery time is\\ $\inf\{t:\delta_t\le d(t,m^\ast/X_t)\}\wedge T$.
\end{enumerate}
\end{theorem}
\begin{proof}
(i) The pair $(X,\delta)$ is explicit: $\delta_s=\delta e^{-\kappa(s-t)}+\alpha(1-e^{-\kappa(s-t)})+\sigma_\delta\int_t^se^{-\kappa(s-u)}\dd W^2_u$ and $X_s=x\hat X_s$ with $\hat X_s=\exp(\int_t^s(r-\delta_u)\dd u-\tfrac{\sigma^2}2(s-t)+\sigma(W^1_s-W^1_t))$. Integration by parts in the OU stochastic convolution bounds $\sup_{t\le s\le T}|\delta_s|$ by $|\delta|+|\alpha|+2\sigma_\delta\sup_{t\le s\le T}|W^2_s-W^2_t|$; the formula for $\hat X$ then bounds $\sup_{t\le s\le T}|\log\hat X_s|$ by a constant plus a linear combination of the two Brownian suprema. These Gaussian suprema have finite exponential moments of every linear order, so $\E\sup_{t\le s\le T}(\hat X_s^p+|\delta_s|^p)<\infty$ for every $p\ge1$. Thus the class (D) property holds, $A$ has integrable total variation, and the martingale in the discounted-price decomposition is square-integrable, verifying Assumption~\ref{ass:semimart}. The scaling is the linearity of the payoff in $x$. Continuity in $x$ is as in Theorem~\ref{thm:gbm}; in $\delta$, $\delta\mapsto\hat X$ is continuous in $L^1(\sup)$ by dominated convergence (the map $\delta\mapsto\delta_\cdot$ is affine); in $t$, as in Theorem~\ref{thm:gbm} by time-homogeneity.

(ii) The slope argument of Lemma~\ref{lem:monotone-m} gives the monotonicity in $y$. For fixed $\tau$ and $\delta<\delta'$, the explicit formulas give $\delta_s\le\delta'_s$ and hence $\hat X_s\ge\hat X'_s$ pathwise, so $\E[e^{-r\tau}(y-\hat X_\tau)]\le\E[e^{-r\tau}(y-\hat X'_\tau)]$; taking suprema, $v$ is nondecreasing in $\delta$, and $y-1$ does not depend on $\delta$. By \ref{F3} the sections of $\cS$ are closed, so $\cS_t=\{\delta\le d(t,y)\}$ with $d(t,y):=\sup\{\delta\in\R:(t,y,\delta)\in\cS\}$, and $\cS_t=\{y\ge k(t,\delta)\}$ with $k(t,\delta):=\inf\{y\in\R:(t,y,\delta)\in\cS\}$. Monotonicity of $d$ in $y$ and of $k$ in $\delta$ follows from the monotonicity of $v-(y-1)$; monotonicity in $t$ from \ref{F2}.

(iii) The running reward is $f(x,\delta)=(\cL-r)(m-x)=\delta x-rm$, i.e.\ $x(\delta-ry)$. Proposition~\ref{prop:nearT}(i) and \ref{F2} give $(t,y,\delta)\in\cC$ for all $t<T$ when $\delta>ry$, so $d(t,y)\le ry$ for every $y\in\R$. This also gives $k(t,\delta)\ge\delta/r> -\infty$. Suppose $d(T-,y_0):=\lim_{t\uparrow T}d(t,y_0)<ry_0$ for some $y_0\in\R$, and pick $\delta_0\in(d(T-,y_0),ry_0)$. Take $x_0=1$ and $m=y_0$. Then $f<0$ on a neighbourhood of $(x_0,\delta_0)=(1,\delta_0)$, so by Proposition~\ref{prop:nearT}(ii) $(t,y_0,\delta_0)\in\cS$ for $t$ close to $T$, i.e.\ $\delta_0\le d(t,y_0)\le d(T-,y_0)<\delta_0$, a contradiction.

(iv) By (i), $V(0,X_0,\delta_0;m)=0$ if and only if $v(0,m/X_0,\delta_0)=0$. The function $v(0,\cdot,\delta_0)$ is strictly increasing (Proposition~\ref{prop:dinkelbach}), with $v(0,0,\delta_0)=-\inf_\tau\E[e^{-r\tau}\hat X_\tau]\le-\E[\inf_{s\le T}e^{-rs}\hat X_s]<0$ and $v(0,1,\delta_0)\ge0$ (take $\tau=0$), so it has a unique zero $k^\ast(\delta_0)\in(0,1]$, and $m^\ast=X_0k^\ast(\delta_0)$. Since $v$ is nondecreasing in $\delta$, the zero $k^\ast$ is nonincreasing in $\delta_0$, and so is $c^\ast$. Next, $c^\ast(\delta_0)=1$ if and only if $v(0,1,\delta_0)=0$ if and only if $(0,1,\delta_0)\in\cS$ if and only if $\delta_0\le d(0,1)$. By (iii), $d(0,1)\le r$, and $d(0,1)$, as a function of the window length $T$, is $d(T-s,1)|_{s=T}$, nonincreasing in $T$ by \ref{F2} and tending to $r$ as $T\downarrow0$.

(v) Lemma~\ref{lem:monotone-m} with $\cS_m=\{y\ge k\}$, $y=m/x$: $(t,x,\delta)\in\cS_m$ if and only if $m\ge x\,k(t,\delta)$.
\end{proof}

The numerical grid in $\log y$ used below resolves only the positive portion of this signed index; the representation \eqref{eq:rep} uses the full signed signal $K$, without truncation at zero.

\section{Numerical results}\label{sec:numerics}

The Python code, saved numerical results, and instructions for reproducing the figures and tables are available at \url{https://github.com/ebayr/eom-gittins}.

All computations use finite-difference schemes with exact projection on the obstacle, run on several grid levels; details and the scripts are described in Appendix~\ref{app:num}. Boundaries are located by fitting the quadratic vanishing of $V-g$ at the boundary to the first two continuation nodes, which removes the plateaus that a node-based reading of the boundary produces. Grid refinement and selected cross-checks are reported below; the displayed values are numerical estimates.

\subsection{Geometric Brownian motion}
\begin{figure}[t]\centering\includegraphics[width=0.72\textwidth]{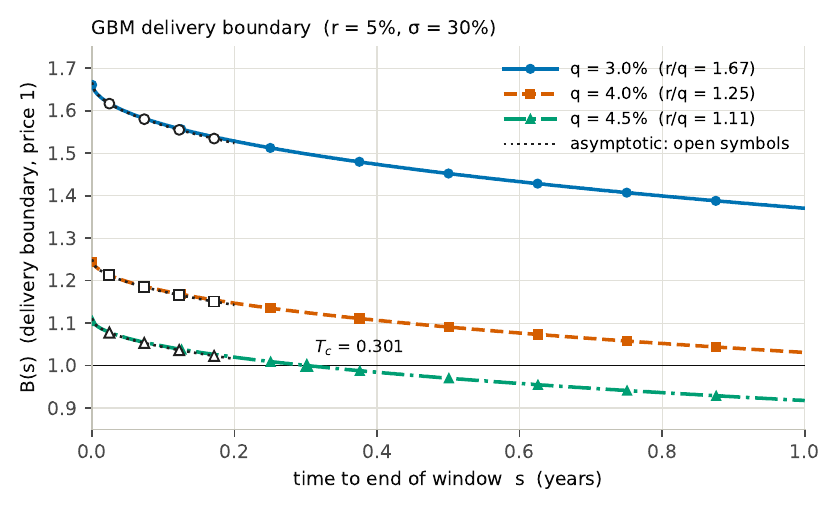}
\caption{Boundary $B(s)$ of the delivery region (price $1$) for $r=5\%$, $\sigma=30\%$; it starts at $r/q$ and crosses $1$ at $T_c$. Filled circles, squares and triangles identify the three numerical curves. Black dotted curves with matching open symbols show the square-root asymptotic of Theorem~\ref{thm:parabolic} on $0\le s\le0.2$ years.}\label{fig:boundary}\end{figure}

\begin{figure}[t]\centering\includegraphics[width=0.72\textwidth]{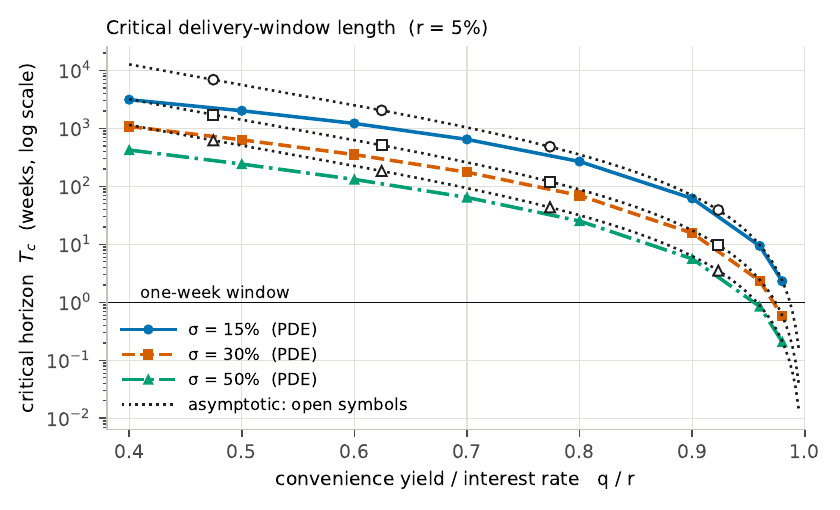}
\caption{Critical horizon $T_c$ (weeks, log scale) against $q/r$ for $r=5\%$. Below a numerical curve $m^\ast=X_0$. Filled circles, squares and triangles identify the three volatility values. Black dotted curves with matching open symbols show $((r/q-1)/(C\sigma))^2$.}\label{fig:phase}\end{figure}

Figure~\ref{fig:boundary} shows the boundary $B(s)$ of Theorem~\ref{thm:gbm} for $r=5\%$, $\sigma=30\%$ and several $q<r$, and Figure~\ref{fig:phase} the critical horizon $T_c$. Table~\ref{tab:Tc} lists $T_c$. For a one-week window with $\sigma=30\%$ the price equals spot unless $q>0.974\,r$ (by Corollary~\ref{cor:Tc}, $C\sigma\sqrt{1/52}=0.0266$); for the sampled one-week cases $q\ge5.5\%$, $c^\ast$ differs from the terminal-delivery price ratio $e^{(r-q)T}$ by less than $0.01$~bp of spot; at $q=r$ the discount is $0.085$~bp; even for a one-\emph{year} window with $q=4.5\%$ the discount $1-c^\ast$ is $6.9$~bp. The constant-yield model thus makes Bj\"ork's problem a bet on the sign of $r-q$ with an $O(\sigma\sqrt T)$ fuzz around $q=r$.

\begin{table}[t]\centering\small
\begin{tabular}{@{}lrrrrrrrr@{}}\toprule
$q/r$ & 0.98 & 0.96 & 0.90 & 0.80 & 0.70 & 0.60 & 0.50 & 0.40\\\midrule
$T_c$, $\sigma=15\%$ (years) & 0.0444 & 0.181 & 1.195 & 5.17 & 12.4 & 23.4 & 38.9 & 60.3\\
$T_c$, $\sigma=30\%$ (years) & 0.0111 & 0.0454 & 0.301 & 1.342 & 3.38 & 6.81 & 12.2 & 20.8\\
$T_c$, $\sigma=50\%$ (years) & 0.0040 & 0.0163 & 0.109 & 0.487 & 1.24 & 2.54 & 4.67 & 8.17\\
$((r/q-1)/(C\sigma))^2$, $\sigma=30\%$ & 0.0113 & 0.0473 & 0.336 & 1.70 & 5.00 & 12.1 & 27.2 & 61.3\\\bottomrule
\end{tabular}
\caption{Critical horizon for $r=5\%$, from the finest of three grid levels ($6401$ price nodes); selected refinement results are discussed below. The last row is the asymptotic formula of Corollary~\ref{cor:Tc}, within $2\%$ at $q/r=0.98$ and $4\%$ at $q/r=0.96$.}\label{tab:Tc}\end{table}

\emph{The square-root asymptotic.} The boundaries of Figure~\ref{fig:boundary} follow $(r/q)(1-C\sigma\sqrt s)$ with implied constants $0.62$--$0.65$ at $s=0.0025,0.01,0.04,0.09$, illustrating Theorem~\ref{thm:parabolic}, and the last row of Table~\ref{tab:Tc} is the formula of Corollary~\ref{cor:Tc} in the form $((r/q-1)/(C\sigma))^2$. The constant $C=0.638833$ is the root of the equation of Lemma~\ref{lem:brownian} with the integrals evaluated by quadrature; a direct finite-difference solve of the Brownian problem \eqref{eq:wdef} gives $0.638$--$0.640$. The accuracy of $T_c$ deserves a comment: reading the boundary off the grid as the last node in the stopping set makes $B$ equal to $1$ exactly over a range of maturities, so that ``the first $s$ with $B(s)\le1$'' and ``the first $s$ with $B(s)<1$'' bracket $T_c$ from below and above with a gap that closes only like the node spacing; at $q/r=0.98$, $\sigma=30\%$, the brackets are $[0.0091,0.0128]$, $[0.0101,0.0120]$ and $[0.0106,0.0116]$ on the three grids, while the fitted boundary gives $0.01104$, $0.01110$ and $0.01111$.

\paragraph{Verification of Theorem~\ref{thm:signal}.}
For geometric Brownian motion with $r=5\%$, $q=4.5\%$, $\sigma=30\%$, $T=1$ (so that $T>T_c=0.30$), the boundary from the implicit scheme agrees with an explicit scheme to four digits and with a $3000$-step binomial tree at the level of the value; $c^\ast=0.999312$ and $B(1)=0.918$. A Monte Carlo simulation with $4\times10^5$ paths of the rule $\tau^{m^\ast}=\inf\{0\le t<T:X_t\le m^\ast B(T-t)\}\wedge T$ gives $\E[e^{-r\tau}(m^\ast-X_\tau)]=0.0004\pm0.0003$, the ratio $\E[e^{-r\tau}X_\tau]/\E[e^{-r\tau}]=0.99894$, and both sides of \eqref{eq:fixed} equal to $0.0054$ to within $10^{-4}$. A single Newton step \eqref{eq:newton} from $m_0=X_0$, computed by solving the two linear equations for $\E[e^{-r\tau_0}X_{\tau_0}]$ and $\E[e^{-r\tau_0}]$ with $\tau_0=\inf\{0\le t<T:X_t\le B(T-t)\}\wedge T$, gives $m_1=0.999312=c^\ast$ to six digits: the observed contraction factor is $7\times10^{-4}$, far below the guaranteed $1-e^{-rT}=0.049$, as expected from the superlinear convergence of Theorem~\ref{thm:newton}.

\subsection{Gibson--Schwartz}
\begin{figure}[t]\centering\includegraphics[width=0.95\textwidth]{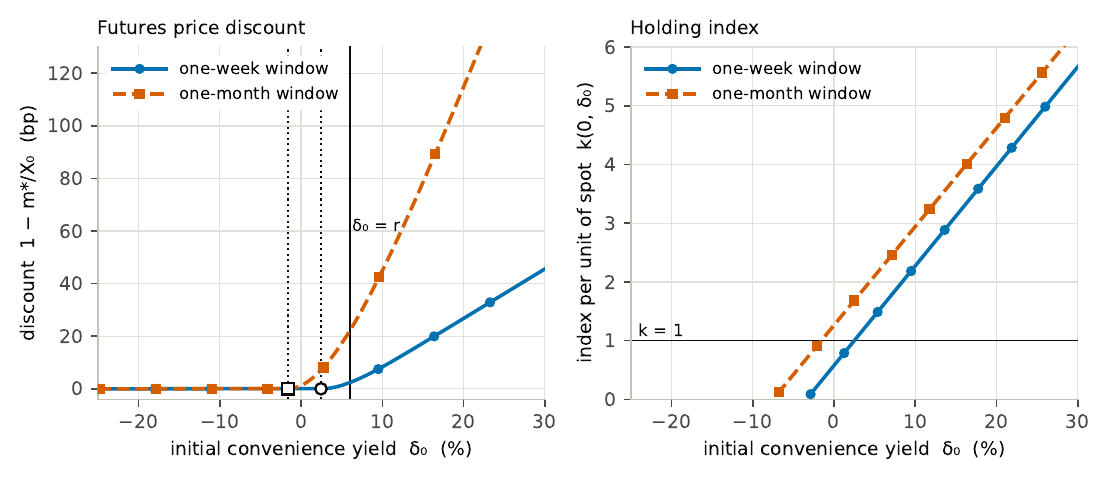}
\caption{Gibson--Schwartz. Solid lines with circles: one week; dashed lines with squares: one month. Left: discount $1-m^\ast/X_0$, rising as the price falls with convenience yield; dotted lines mark $\delta_c$, with matching open symbols at zero discount. Right: normalized index $k(0,\delta_0)=K(0,X_0,\delta_0)/X_0$ where resolved by the grid, to be compared with $c^\ast=m^\ast/X_0$. The line $k=1$ marks the initial immediate-delivery threshold.}\label{fig:gs}\end{figure}

\begin{table}[t]\centering\small
\begin{tabular}{@{}lrrrrrrrr@{}}\toprule
$\delta_0$ & $-20\%$ & $0\%$ & $2\%$ & $3\%$ & $4\%$ & $6\%$ & $10\%$ & $20\%$\\\midrule
one week: $1-c^\ast$ (bp) & 0.00 & 0.00 & 0.00 & 0.07 & 0.51 & 2.36 & 8.33 & 26.8\\
\quad frozen-yield carry $1-e^{(r-\delta_0)T}$ (bp) & $-50.1$ & $-11.5$ & $-7.7$ & $-5.8$ & $-3.8$ & 0.0 & 7.7 & 26.9\\
\quad index $k(0,\delta_0)$ & $<0.05$ & 0.59 & 0.92 & 1.09 & 1.26 & 1.60 & 2.28 & 3.97\\
one month: $1-c^\ast$ (bp) & 0.00 & 1.22 & 5.72 & 9.00 & 12.9 & 22.1 & 44.9 & 115\\
\quad index $k(0,\delta_0)$ & $<0.05$ & 1.27 & 1.61 & 1.77 & 1.94 & 2.28 & 2.95 & 4.63\\\bottomrule
\end{tabular}
\caption{Gibson--Schwartz (oil parameters, $r=6\%$): discount of the futures price below spot and the holding index per unit of spot, $k(0,\delta_0)$, at the yields shown (which are grid nodes); $c^\ast=1$ exactly when $k\le1$. Discounts are from the finest grid, indices are Richardson-extrapolated from the two finest grids; ``$<0.05$'' means that the stopping region reaches the lower edge $y=e^{-3}\approx0.05$ of the computational domain.}\label{tab:gs}\end{table}
Holding $X_0$ fixed, a higher initial convenience yield lowers future spot prices under the same pricing-measure shocks. This raises the short's payoff $m-X_\tau$ and hence its value at fixed $m$, so the zero-value condition requires a weakly lower $m^\ast$. Figure~\ref{fig:gs} therefore shows a rising discount $1-m^\ast/X_0$, remaining zero for $\delta_0\le\delta_c(T)$.

The holding index is compared with the fixed delivery price: for $t<T$, delivery is optimal when $K(t,X_t,\delta_t)=X_tk(t,\delta_t)\le m^\ast$, or equivalently $k(t,\delta_t)\le m^\ast/X_t$. At inception, Theorem~\ref{thm:gs} gives two cases: $k(0,\delta_0)\le1$ and $m^\ast/X_0=1$ (immediate delivery), or $k(0,\delta_0)>1>m^\ast/X_0$ (continuation). This explains the horizontal line at $1$ in the right panel; later the normalized delivery threshold is $m^\ast/X_t$.

With the oil parameters of \cite{Schwartz97} ($\kappa=1.876$, $\alpha=0.106$, $\sigma=0.393$, $\sigma_\delta=0.371$, $\rho=0.766$, $r=6\%$) the two-dimensional obstacle problem for $v(t,y,\delta)$, solved by a finite-difference scheme with a monotone interior stencil on four grid levels (Appendix~\ref{app:num}), gives Table~\ref{tab:gs} and Figure~\ref{fig:gs}. The critical yield of Theorem~\ref{thm:gs}(iv) is $\delta_c=2.5\%$ for a one-week window and $-1.6\%$ for a month, against $r=6\%$ (the four grid levels give $2.75$, $2.44$, $2.45$, $2.46\%$ and $-1.4$, $-1.6$, $-1.6$, $-1.6\%$; an independent estimate from the root of $k(0,\cdot)=1$ gives $2.50\%$ and $-1.55\%$ on the finest grid). The heuristic analogue of Theorem~\ref{thm:parabolic} for this model (Remark~\ref{rem:parabolic}(a), with the volatility of the convenience flow $\delta X$ in place of $\sigma\bar x$), $\delta_c+C\sqrt T\,\big(\sigma_\delta^2+2\rho\sigma\sigma_\delta\delta_c+\sigma^2\delta_c^2\big)^{1/2}=r$, gives $2.6\%$ and $-0.8\%$; we have not proved it, and it is only an asymptotic statement. At $\delta_0=r$ the discount is $2.4$~bp of spot over a week (about thirty times the constant-yield value at $q=r$, since it now lives on the volatility of $\delta$ rather than that of the price) and $22$~bp over a month. At $\delta_0=20\%$, the computed discounts are close to the frozen-yield benchmark $1-e^{(r-\delta_0)T}$; this benchmark omits mean reversion and yield uncertainty.

\section{Concluding remarks}\label{sec:conclusion}

Bj\"ork asked for the price of a contract that is a ratio of two expectations and for the delivery strategy behind it. The answer is that the contract is a bandit: the short holds a project whose reward is the convenience yield of the underlying---a proportional flow in the constant-yield model and a stochastic flow in the Gibson--Schwartz model---and retires by delivering. The futures price is the retirement reward at which the project is worth the spot, delivery is optimal the first time the project's Gittins index falls to the price, and the index is the Bank--El Karoui signal of the excess of the discounted price over its terminal expectation; Dinkelbach's iteration computes the price from standard stopping problems with a guaranteed rate.

A remaining question is the robust version of the problem under drift uncertainty. Dinkelbach's reduction survives: the short's price is the zero of $m\mapsto\sup_\tau\sup_{\P\in\mathcal P}\E^\P[D_\tau(m-X_\tau)]$, the long's that of the $\inf_{\P}$ version, and the fair price becomes an interval, with Theorem~\ref{thm:newton} available for the short's end. The representation theorem and hence the index do not survive as such: under a sublinear expectation the signal would have to be an essential supremum over stopping times \emph{and} priors, and the level-crossing structure, which rests on the linearity of the representation, is not available. This calls for a different set of tools and is the subject of a separate paper.

\appendix
\section{Numerical methods}\label{app:num}
\begin{itemize}[leftmargin=1.2em,itemsep=3pt]
\item \emph{Geometric Brownian motion.} Implicit Euler in log-price with the Brennan--Schwartz projected tridiagonal solve (exact for left-interval stopping regions), on three grid levels with $1601$, $3201$ and $6401$ nodes and $1000$, $2000$ and $4000$ time steps; checked against an explicit scheme and a CRR tree. The zero $x^\ast(s)$ of $V(s,\cdot)$ is obtained by linear interpolation of the sign change. The boundary $B(s)$ is obtained from the first two continuation nodes $x_j<x_{j+1}$ by fitting $V-g\approx a(x-B)^2$ (smooth fit), $B=(x_{j+1}\sqrt{u_j}-x_j\sqrt{u_{j+1}})/(\sqrt{u_j}-\sqrt{u_{j+1}})$, and $T_c$ by interpolating this $B$ across $1$; the node-based brackets of $T_c$ are reported in Section~\ref{sec:numerics}.
\item \emph{Gibson--Schwartz.} Explicit scheme in $(\log y,\delta)$ with equal spacing $h$ in both variables on $[-3,2]\times[-0.9,0.9]$, $h=0.01,0.005,0.0025,0.00125$ (up to $4001\times1441$ nodes). The mixed derivative is discretized with the seven-point stencil on the anti-diagonal (the cross coefficient is $-\rho\sigma\sigma_\delta<0$ in these variables). The inequalities $\sigma\ge\rho\sigma_\delta$ and $\sigma_\delta\ge\rho\sigma$ for the oil parameters ensure nonnegative off-diagonal diffusion weights; with central differences for the first-order terms, positivity of the interior weights is checked at run time. A time step below the positivity limit and exact obstacle projection make the interior update monotone. Linear extrapolation supplies the edge values; this boundary closure is not monotone. The yields of Table~\ref{tab:gs} are grid nodes. The index $k(0,\delta)$ and the critical yield $d(0,1)$ are obtained from the quadratic fit above in the $\log y$ and $\delta$ directions respectively.
\item \emph{The constant $C$.} Root of the equation of Lemma~\ref{lem:brownian} with the integrals evaluated by quadrature, and a finite-difference solve of the Brownian problem \eqref{eq:wdef} on $[0,1]$. The repository linked in Section~\ref{sec:numerics} contains the scripts and saved outputs; its README gives installation and execution commands.
\end{itemize}

\end{document}